\documentclass{aa}
\usepackage{psfig}
\usepackage{natbib}
\usepackage{lscape}
\usepackage{wasysym}

\usepackage{graphics}
\usepackage{txfonts}
\usepackage{color}
\usepackage{times}

\newfont{\vssn}{cmss10 scaled 1050}
\newfont{\vsss}{cmss10 scaled 450}
\newfont{\nsf}{cmssdc10 scaled 1000}
\newfont{\vs}{cmssdc10 scaled 700}
\newfont{\lvss}{cmssdc10 scaled 900}
\newfont{\lx}{cmssdc10 scaled 760}
\newfont{\llx}{cmssdc10 scaled 1000}
\newfont{\nlx}{cmssdc10 scaled 900}

\def\ko{K\"O00}

\def\hst{{\sl HST}}

\def\H1{\ion{H}{i}}
\def\p25{{\it P$_{25}$}}
\def\e25{{\it E$_{25}$}}
\def\reff{{\it R$_{\rm eff}$}}
\def\rsf{{\it R$_{\rm SF}$}}

\newcommand{\kmsec}{km s$^{-1}$}
\newcommand{\msun}{$M_\odot$}

\newcommand{\sbb}{mag/$\sq\arcsec$}
\newcommand{\oh}{12+log(O/H)}
\def\ha{H$\alpha$}
\def\hb{H$\beta$}
\def\o5007{[O {\sc iii}] $\lambda$5007}
\def\h2{H{\sc ii}}
\def\cc{\object{I~Zw~18\,C}}
\def\iz18{\object{I~Zw~18}}
\def\rr{{\sl R}$^{\star}$}
\def\P25{{\sl R}$_{\rm SF}$}
\def\E25{{\sl R}$_{\rm host}$}
\def\eqan{\begin{equation}}
\def\eqen{\end{equation}}
\def\sfha{\lvss SFH1\rm}
\def\sfhb{\lvss SFH2\rm}
\def\sfhc{\lvss SFH3\rm}
\def\sfhe{\lvss SFH2,3\rm}
\def\ige{\nsf ne\rm}
\def\cps{\nsf cps\rm}
\def\dc{$\delta_{\rm ce}$}
\def\ewhn{EW(H$\alpha$+[N\,{\sc ii}])}

\newcounter{qub}
\begin{document}
%
\title{I\ Zw\ 18 as morphological paradigm for rapidly assembling high-$z$ galaxies
}
\author{P. Papaderos \inst{1,2}
\and G. \"Ostlin\inst{2}
}
\offprints{P. Papaderos; papaderos@astro.up.pt}
\institute{Centro de Astrof{\'\i}sica and Faculdade de Ci\^encias, Universidade do Porto,
Rua das Estrelas, 4150-762 Porto, Portugal
\and
Department of Astronomy,
Oskar Klein Centre,
Stockholm University,
SE - 106 91 Stockholm,
Sweden
}
\date{Received \hskip 2cm; Accepted}

\abstract
{\iz18, ever since regarded as the prototypical blue compact dwarf (BCD)
galaxy, is, quite ironically, the most atypical BCD known.
This is because its large low-surface brightness (LSB) envelope is not due 
to an old underlying stellar host, as invariably is the case for typical BCDs, 
but entirely due to extended nebular emission (Papaderos et al. 2002; hereafter P02).}
{Our goal is to explore \iz18\ and its detached C~component \cc\ down to an
unprecedently faint surface brightness $\mu$ (\sbb) level in order to gain further 
insight into the structural properties and evolutionary history of this enigmatic galaxy pair.
}
{We present a photometric analysis of the entire set of archival 
\hst\ ACS $V$, $R$ and $I$ band data for \iz18.
}
{Radial color profiles for \cc\ reveal blue and practically 
constant colors (0$\pm$0.05) down to $\mu\!\!\sim\!\!27.6$, and a 
previously undisclosed, slightly redder ($V$--$I$$\approx$0.2), 
stellar population in its extreme periphery ($\mu\sim$29).
We argue that stellar diffusion over $\tau\!\!\sim\!\!10^8$ yr and the associated
\emph{stellar mass filtering effect} (P02) can consistently account 
for the observed properties of the stellar component in the outskirts of \cc.
This process, in combination with propagating star formation with a mean
velocity of $\sim$20 \kmsec\ can reproduce all essential characteristics of 
\cc\ within $\sim\tau$.
An extremely faint substrate of older stars can neither be ruled out nor does
need be postulated.
As for \iz18, we find that nebular emission (\ige) extends out to 
$\sim$16 stellar scale lengths, shows a nearly exponential outer profile, 
and provides at least 1/3 of the total optical emission. 
\ige\ dominates already at $\mu \sim$23.5, as evident from e.g. the uniform 
and extremely blue ($V$--$I$$\approx$--1, $R$--$I$$\approx$--1.4) 
colors of the LSB envelope of \iz18.  
}
{The case of \iz18\ suggests caution in studies of distant galaxies in 
dominant stages of their evolution, rapidly assembling their stellar mass 
at high specific star formation rates (SSFRs).
It calls attention to the fact that \ige\ is not necessarily cospatial with the 
underlying ionizing and non-ionizing stellar background, neither has to scale 
with its surface density. The prodigious energetic output during dominant phases 
of galaxy evolution may result in large \emph{exponential} \ige\ envelopes,
extending much beyond the still compact stellar component, just like in \iz18.
Therefore, the morphological paradigm of \iz18, while probably unique in the nearby Universe, 
may be ubiquitous among high-SSFR galaxies at high redshift. 
Using \iz18\ as reference, we show that extended \ige\ may introduce
substantial observational biases and affect several of the commonly studied fundamental galaxy relations.
Among others, we show that the surface brightness profiles of distant morphological 
analogs to \iz18\ may be barely distinguishable from S\'ersic profiles with an 
exponent 2$\la\eta\la$5, thus mimicking the profiles of massive galaxy spheroids. 
}

\maketitle

\markboth {Papaderos \& \"Ostlin}{I\ Zw\ 18 as morphological paradigm for rapidly assembling high-$z$ galaxies}

\section{Introduction \label{intro}}
Even four decades after its discovery \citep{SS70}, the blue compact dwarf
(BCD) galaxy \iz18\ continues to attract considerable interest and feed 
intense debates in extragalactic research. 
Its low oxygen abundance \citep{SS72}, established in numerous subsequent studies
\citep[][among others]{Lequeux79,French80,KD81,Pagel92,SK93,Martin96,VilchezIglesiasParamo98-IZw18,IT98a,IT98b,ICF99} to be \oh$\approx$7.2, makes it the third most metal-poor star-forming (SF) 
galaxy in the nearby Universe, after \object{SBS\ 0335-052\,W} 
\citep{Izotov05-SBS0335,Papaderos06-SBS0335,Izotov09-SBS0335} and
\object{DDO68} \citep{Pustilnik05-DDO68,IT07-DDO68}.  
Despite a meanwhile long record of extremely metal-poor (\oh$\la$7.6) BCDs 
(hereafter XBCDs) discovered in the recent years 
\citep[see e.g.][for a review]{Papaderos08,Guseva09-LZ},
\iz18\ remains the unconquered prototypical example of this enigmatic
galaxy class.

\begin{figure*}[ht]
\begin{picture}(17.4,16.4)
\put(0.2,0){{\psfig{figure=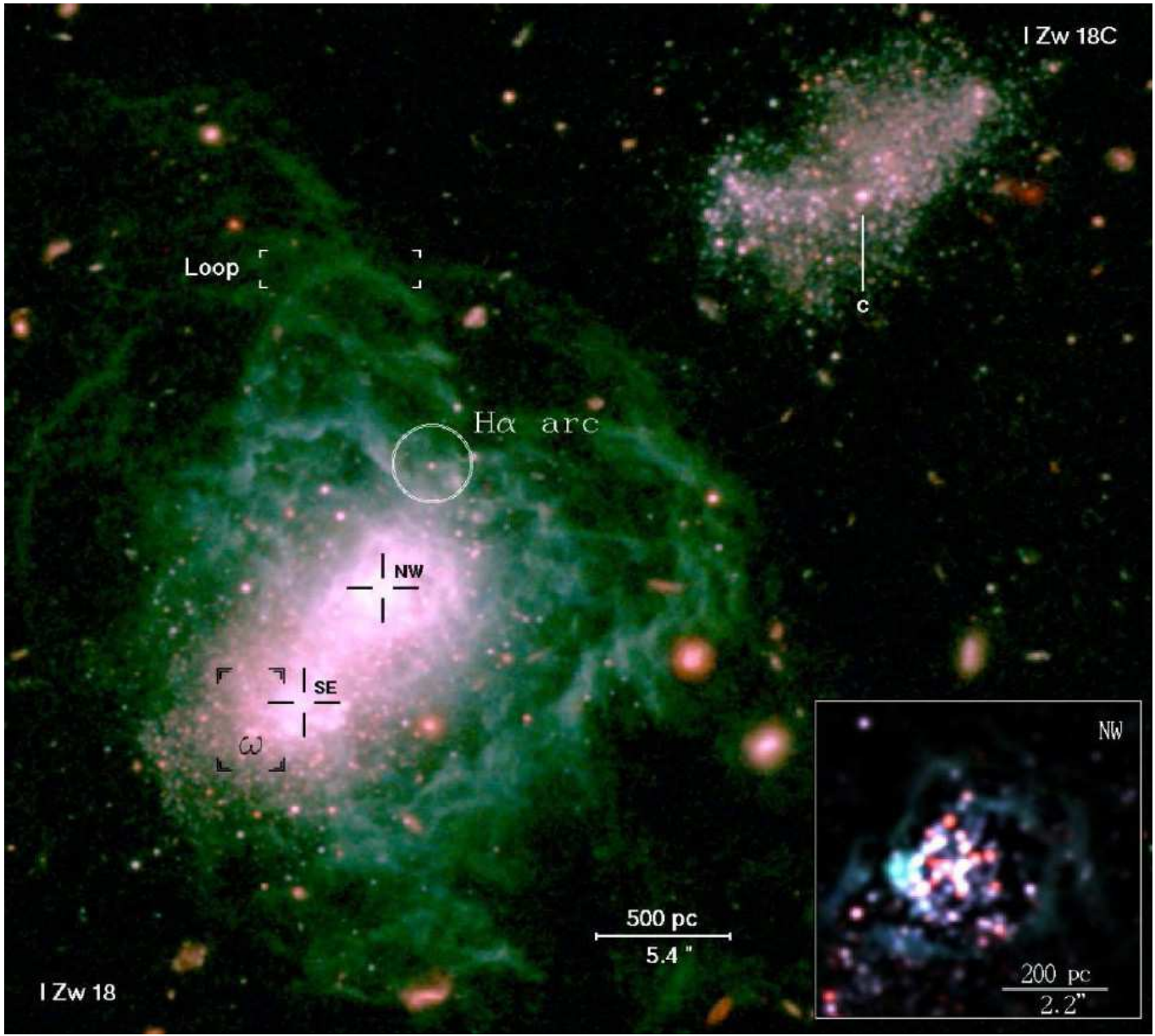,width=18.0cm,angle=0,clip=}}}
\end{picture}
\caption[]{
Three-color composite image of \iz18\ and \cc, combining 
\hst\ ACS data in $V$, $R$ and $I$ (blue, green and red channel,
respectively). The position of the star-forming regions NW and SE 
is indicated by crosses. The regions labeled ``Loop'' and ``H$\alpha$ arc'' 
were studied through deep Keck\,II long slit spectroscopy by \cite{Izotov01-IZw18}. 
The region $\omega$ at the southeastern tip of \iz18\ shows comparatively weak
nebular emission, its colors allow therefore to place meaningful constraints 
on the age of the stellar component \citep{Papaderos02-IZw18}.
In the magnified version of region NW (inset), combining the unsharp masked 
images $I_{\rm c}$, $R_{\rm c}$ and $V_{\rm c}$ (see Sect. \ref{results} 
for details), about 30 point sources, surrounded by a complex network 
of ionized gas shells are discernible. The irregular blue galaxy \cc\ 
is located $\sim$22\arcsec\ northwest of region NW ($\approx$2 kpc at 
the assumed distance of 19 Mpc to \iz18). It shows faint nebular 
emission in its bluer southeastern tip and central star cluster 
complex C \citep{Izotov01-IZw18}. North is at the top and east to the left.
}
\label{IZw18-Fig1}
\end{figure*}

\iz18\ was originally described by \cite{Zwicky66} as a pair of compact galaxies, 
later on recognized to be SF regions within the same galaxy, the brighter
northwestern (NW) and the fainter southeastern (SE) component, separated by
$\approx$6\arcsec\ (cf Fig. \ref{IZw18-Fig1}).  
Subsequent work has shown that these regions are embedded within an extended,  
low-surface brightness (LSB) envelope
\citep{Davidson89,DufourHester90,Ostlin96,Dufour96a,Martin96},
whose rich filamentary substructure was impressively revealed with the 
advent of the \hst\ \citep{HT95,Dufour96b}.
That nebular emission (hereafter \ige) is very strong in the central part of 
\iz18\ and its north-western super-shell was spectroscopically documented early on.
For example, \cite{Izotov01-IZw18} have shown on the basis of deep 
long slit spectroscopy that the equivalent width (EW) of the \ha\ emission 
line rises to 1700~$\AA$ northwest of region NW and that \ige\
is present as far away as 15\arcsec\ from it (regions labeled ``\ha\ arc''
and ``Loop'' in Fig. \ref{IZw18-Fig1}). 
The EW(\ha) morphology of \iz18\ was first studied 
with high-resolution ground-based imagery by \cite{Ostlin96} who
described a horseshoe-shaped rim of intense (EW(\ha)$\simeq$1500 $\AA$) \ige\ 
encompassing region NW. This conspicuous EW pattern was later on confirmed through 
\hst\ WFPC2 data \citep{Papaderos01-IZw18,Izotov01-IZw18} and, more
impressively, by \cite{VilchezIglesiasParamo98-IZw18} who were the first to 
present a 2D spectroscopic study of the chemical abundance patterns 
of the warm interstellar medium (ISM) in \iz18.

Much less attention has been attracted by the fainter detached C component of
\iz18\ (hereafter \cc), located $\sim$ 22\arcsec\ northwest of region NW. 
\cite{Dufour96a}, \cite{Petrosian97}, \cite{IT98a}, \cite{vanZee98-IZw18} and \cite{Izotov01-IZw18}
have shown it to have the same recession velocity as the main body, thus establishing its 
physical association to \iz18. This was also shown through interferometric 21cm studies
\citep[][see also \cite{Via87}]{vanZee98-IZw18} which revealed that \iz18\ and \cc\ are 
immersed within a large common HI complex with a projected size of 60\arcsec $\times$
45\arcsec\ connecting with a $\ga$1\arcmin\ southern tail with no optical counterpart.
The SF activity in \cc\ is known to be weak with its EW(\ha) not exceeding $\sim$60 $\AA$ 
along its major axis \citep[][see also \citet{vanZee98-IZw18}]{Izotov01-IZw18}.
Despite deep Keck\,II spectroscopy, \cite{Izotov01-IZw18} failed to detect
oxygen lines, so its oxygen abundance is not known.

Traditionally, the distance to \iz18\ has been taken to be 10 Mpc, assuming a
pure Hubble flow recessional velocity. However, \cite{Izotov99-IZw18} have argued
based on an \hst\ color-magnitude diagram (CMD) study that the distance to
\iz18\ has to be at least 15 Mpc, and most likely $\sim$20 Mpc, in order for
its brightest stars being massive enough to account for the ionizing flux observed.
This upper distance value has recently received independent support by 
\cite{Fiorentino10}. These authors have identified three long-period Cepheid 
candidates in \iz18, which, if interpreted as classical Cepheids, imply by the 
Wesenheit relation a distance of 19.0$^{+2.8}_{-2.5}$ Myr. 
In the following, we shall throughout adopt a distance $D$=19 Mpc to both \iz18\ and
\cc\ and convert distance-dependent quantities from the literature
accordingly. It should be noted, however, that the assumed distance has practically 
no influence on the main conclusions from this study.

The wealth of dedicated studies of \iz18\ highlight the importance placed on
this XBCD as precious nearby laboratory for exploring collective star
formation and the associated feedback process under metallicity conditions 
approaching those in distant protogalactic systems. 
Some examples include the consideration of \iz18\ as reference object for many
dwarf galaxy chemical evolution models
\citep[][among others]{MatteuchiTosi85,RoyKunth95,MHK99,Legrand00-IZw18,Recchi04-IZw18}, 
the great deal of effort put in the determination of its chemical abundance
patterns in its neutral ISM \citep{Kunth94-IZw18,Aloisi03-H2,LdE04-IZw18},
and in the study of its dust and molecular gas content 
\citep[e.g.][]{Cannon01,Leroy07-IZw18},
the thorough exploration of the excitation mechanisms of its brightest H{\sc ii} region
\citep{StasinskaScharer99-IZw18,Pequignot08-IZw18}, and the deep spectroscopic 
studies that led to the discovery of Wolf-Rayet stellar features in it 
\citep{Izotov97-IZw18-WR,Legrand97-IZw18-WR}.

But arguably, the most longstanding debate associated with \iz18\ ever since 
its discovery concerns its evolutionary status. 
In this regard, various interpretations have been put forward, ranging 
from \iz18\ being a \emph{bona fide} young galaxy, currently forming 
its \emph{first} stellar 
generation \cite[][see also \citet{IzotovThuan99-heavy-elements}]{SS70},
all through the diametrically opposite picture of an ancient 
``slowly cooking'' dwarf galaxy that is forming stars continuously over the 
Hubble time \citep{Legrand00-IZw18,Legrand01-IZw18}.
Notwithstanding an impressive amount of high-quality multiwavelength data 
and numerous dedicated analyses of considerable effort and sophistication,
the convergence towards a consensual view on the evolutionary status of 
\iz18\ has been slow. 

CMD analyses, based on \hst\ data, have primarily been focusing on the
question of whether or not \iz18\ contains a sizeable population of evolved 
red giant branch (RGB) 
stars, similar to typical \citep[\oh$\ga$8, see e.g.][]{KunthOstlin00} BCDs. 
In the latter, an extended envelope of resolved RGB stars around the 
SF component \citep[e.g.][]{Tosi01,Crone02} nicely echoes the since-long
observationally established fact of an evolved underlying host galaxy in 
these systems \citep[e.g.][]{LT86,P96a,LM01a,BergvallOstlin02,Noeske03-NIR,GildePazMadore05}.
Initial CMD studies suggested for the main body an age between several 
10 Myr \citep{HT95,Dufour96b} and $\sim$~1 Gyr \citep{Aloisi99-IZw18}. 
\citet{Ostlin00} argued from an \hst\ NICMOS near infrared (NIR) 
study that a fit to the $J$ vs. $J-H$ CMD of \iz18\ is best achieved for 
a stellar population of age $\sim$5~Gyr. The subsequent identification of 
five carbon star candidates with an estimated age 0.5--1 Gyr by 
\cite{OstlinMouhcine05} is in accord with that conclusion, even though the 
number of evolved star candidates in all above studies (about a dozen
altogether) was recognized to be surprisingly small compared to typical
BCDs of equal luminosity.
A significant step forward has been possible through the advent of \hst~ACS, 
allowing \cite[][hereafter IT04]{IT04-IZw18} to extend point source photometry
to magnitudes as faint as 29 mag in the $V$ and $I$ band and revisit the
question of the presence of RGB stars in \iz18. 
IT04 found, in addition to numerous blue main sequence and blue and red 
supergiants with an age $\la$100 Myr, an older population of 
asymptotic giant branch (AGB) stars with an age between 0.1 and 0.5 Gyr.
This study, in which no RGB were detected, has been the first to also explore 
the spatial distribution of stars of different ages in \iz18. 
The upper age limit of 0.5 Gyr for the oldest stars in \iz18\ (IT04) was
subsequently relaxed from a re-analysis of the same data by 
\citet{YakobchukIzotov06} which revealed an untypically small number 
of RGB candidates. Various other efforts have been made to improve on the 
CMD analysis of IT04 by pushing point source photometry to 
by 1--2 mag fainter levels \citep{Momany05,Tosi07-IZw18,Tikhonov07-IZw18}.
The faintest ($>$29 mag) point sources in those CMDs cover almost uniformly
the color range between $<$--1 mag and $>$2 mag.
It is worth pointing out that, whereas divergent in their conclusions 
regarding stellar age, all CMD analyses for \iz18\ consistently indicate 
a conspicuous absence of an extended stellar LSB envelope surrounding 
regions NW\&SE, at sharp contrast to any previously studied BCD.

As for \cc, CMD analyses yield an upper age between a few ten and hundred Myr 
\citep{Dufour96b,Aloisi99-IZw18}. Recently, \citet{Jamet10-IZw18} employing a probabilistic
CMD modeling technique reported an upper age of $\sim$125 Myr, without, however,
strictly ruling out the presence of older stars.
An age of the same order was previously inferred for \cc\ from a combined CMD 
and evolutionary spectral synthesis study by \cite{Izotov01-IZw18}.

From the viewpoint of surface photometry, diametrically different
conclusions on the photometric structure and evolutionary status of 
\iz18\ were drawn by \citet[][hereafter \ko]{KunthOstlin00} and 
\citet[][hereafter P02]{Papaderos02-IZw18}. 
Nevertheless, these two studies were the first to demonstrate on the basis of 
surface photometry that \iz18\ is not presently forming its first stellar 
generation but contains a substantial unresolved stellar background of
intermediate age. 

As a matter of fact, much of the disparity between these studies has been due 
to the different importance they ascribed to the presence and photometric 
impact of \ige.
\ko\ concluded that SF activity in \iz18\ is hosted by an old, extended 
stellar disk that dominates the stellar mass, just like in typical BCDs.
Their rationale has mainly been based on their finding that \iz18\ shows 
an exponential intensity decrease and reddish ($B$--$R$$\approx$0.6 mag) 
colors in its LSB envelope (9\arcsec$\ga$\rr$\la$20\arcsec). 
On the assumption that stellar emission dominates throughout, this color
translates by a continuous star formation model to an age of $\ga$5 Gyr.
The central surface brightness $\mu_0$ and exponential scale length 
$\alpha$ of the disk, read off the $B$ band surface brightness profile
(SBP) of \ko, imply that the old stellar host contains $\sim$1/2 of the 
emission and the bulk of the stellar mass in \iz18.
Note that the stellar disk interpretation for \iz18\ is in qualitative agreement 
with the evolutionary scenario by 
\citet[][see also \cite{Legrand00-IZw18}]{Legrand01-IZw18}.
These authors argued that the low and uniform gas-phase 
metallicity of \iz18\ is reproducible through continuous low-level star 
formation throughout the main HI complex of \iz18\ (45\arcsec$\times$60\arcsec)
over the past $\sim$14 Gyr. 
This process would produce an extended stellar disk 
of extremely low surface brightness ($\overline{\mu}\simeq$28 $B$ \sbb).

P02 \citep[see also][]{Papaderos01-IZw18}, on the other hand, called into
question the conclusions by \ko\ by invoking various lines of observational evidence.
First, they have empirically shown that an exponential outer intensity
drop off is a generic property of the nebular halo of starbursting dwarf galaxies.
Consequently, the exponentiality of the LSB envelope of \iz18\ 
is not \emph{per se} a compelling argument for it being due to a stellar disk.
This also applies to its reddish colors which can naturally be accounted for 
by photoionized gas of subsolar metallicity \citep[see e.g.][]{Krueger95}.
Secondly, and in a more straight forward approach, P02 used \hst\ WFPC2 narrow
band images to bidimensionally subtract the [O{\sc iii}] and \ha\ line
emission from broad band \hst\ data in order to isolate and study the residual  
stellar emission in \iz18.
This correction led to the virtual removal of the filamentary LSB envelope,
proving its gaseous nature. Specifically, P02 have shown that \ige\ dominates 
the line-of-sight intensity already at a photometric radius
\rr$\ga$6\arcsec\ and contributes between 30\% and 50\% of the 
$R$ band luminosity of \iz18.
Broad band images, after decontamination from nebular line emission (though still 
affected by nebular continuum emission), were then used 
to study the photometric structure and color distribution of the \emph{stellar} component of \iz18. 
SBPs computed from them have revealed a very compact host which, 
at sharp contrast to typical BCDs, shows practically no radial color 
gradients and overall very blue ($\la$0.1 mag) colors down to a limiting 
surface brightness $\mu\sim$26 \sbb.   
These exceptional properties were interpreted as evidence for youth: 
young stars in \iz18\ did not have had enough time to migrate significantly 
far from heir initial locus and gradually form the extended stellar host 
that is typical of evolved BCDs. This, and the blue optical and NIR colors of the 
southeastern tip of \iz18\ (region $\omega$ in the notation of P02,
cf Fig. \ref{IZw18-Fig1}) where \ige\ is weak led P02 to conclude
that most of the stellar mass in \iz18\ has formed within the past 0.5 Gyr. 
Hence, the picture put forward by P02 is that \iz18\ is a cosmologically 
young object that presently undergoes its dominant formation phase and
contains a small, if any, mass fraction of stars older than $\sim$1 Gyr.
Further support to this conclusion came from a subsequent NIR study of 
\iz18\ by \cite{Hunt03-IZw18}. 
As for \cc, the nearly constant blue colors ($\approx$0 mag) 
determined within its Holmberg radius, lent further support to 
the youth interpretation that was previously advocated by \cite{Izotov01-IZw18}.

However, important aspects of the \iz18\ system could not be conclusively 
addressed from previous photometric studies.
For example, whereas SBPs for \cc\ by P02 reach a surface brightness level 
$\mu\sim28$ $B$ \sbb, their large photometric uncertainties already 
below $\approx$26 \sbb\ have practically prevented an assessment of the question of whether
a redder underlying stellar population dominates in the extremely faint periphery of the galaxy. 
Clearly, this issue is central to the understanding of the evolutionary status of \cc.
One may argue that, since the evolved stellar host generally dominates for 
$\mu\ga24.5$ $B$ \sbb\ \citep[][hereafter P96a]{P96a}, it would have been 
detected in \cc, should have been present.
This is a circular argument, however, given that empirical relations established 
for typical BCDs should not be taken for granted for young dwarf galaxy candidates. 
Similarly, due to the shallowness of previous surface photometry 
no definite conclusions could be drawn regarding the ultra-LSB disk predicted 
by \cite{Legrand00-IZw18}.

As for the main body of \iz18, previous studies did not had the
sensitivity to pin down the maximal extent, morphology and color pattern 
of the LSB envelope, adding potentially important constraints 
to chemodynamical and spectrophotometric models for \iz18.
From such considerations, extremely deep surface photometry appears 
crucially important for further advancing our understanding of the 
photometric structure and evolutionary status of \iz18\ and \cc. 
This is particularly true for deep $I$ band surface photometry 
which is entirely lacking both for \iz18\ and \cc.
Note that $I$ band \hst\ WFPC2 images included the main body of
\iz18\ only and were not deep enough for a study of its LSB envelope.
This data set was therefore not used in the surface photometry analysis by P02.

This study is motivated by the availability of an unprecedently deep set 
of archival \hst\ ACS $V$, $R$ and $I$ broad band data that has accumulated over the 
past few years. $I$ band photometry is an important asset in this respect,
not only due to its higher sensitivity to a putative old stellar
background but also because it offers, in combination with $V$ and $R$ data, 
a sensitive tracer of \ige\ both in the center and the LSB envelope
of \iz18. This is because the $I$ band is affected by nebular continuum emission
only, whereas the $V$ and $R$ transmission curves additionally include, respectively,
the strong [O{\sc iii}]$\lambda\lambda$4959,5007 and \ha\ emission lines.
As a result, regions with strongly enhanced \ige\ can readily be identified by
their extremely blue (--0.5 \dots --1.4) $V$--$I$ and $R$--$I$ colors 
\citep[see e.g.][P02 and references therein for a discussion and 
examples among XBCDs]{Papaderos98-SBS0335}.

This paper is organized as follows: in Sect. \ref{data} we discuss the data
processing and SBP derivation technique used and in Sect. \ref{results} the
structural and morphological properties of \cc\ and \iz18.
Section \ref{discussion} concentrates on the evolutionary status and
the formation process of \cc\ (Sect. \ref{IZw18C-evol}) under consideration of
the effect that the diffusion of young stars would have on the observed
colors (Sect. \ref{mass-filtering}). 
The evolutionary status of \iz18\ and the hypothesis of an ultra-LSB 
underlying stellar disk are discussed on the basis of the 
present photometric analysis in Sects. \ref{age-IZw18} and \ref{izw18-disk}, 
respectively.
In Sect. \ref{iz18-z} we use \iz18\ as template to briefly explore the biases 
that extended \ige\ may introduce in photometric studies of morphologically
analogous star-forming galaxies at higher redshift ($z$).
The main results from this study are summarized in Sect. \ref{Conclusions}.

\section{Data processing \label{data} }
This study is based on the entire set of archival \hst\ ACS broad band images 
for \iz18\ that has been acquired through the \hst\ programs 9400 (PI: Thuan) and 
10586 (PI: Aloisi). It comprises 38, 65 and 81 images in the filters 
F555W ($V$), F606W (broad $VR$, referred to in the following as $R$) 
and F814W ($I$), summing up to on-source exposures of 87, 55 and 101
ksec, respectively. This is the deepest imaging data set currently available 
for \iz18, with an integration time in $R$ and $I$ equaling $\sim$1/3 of 
the time spent on the \hst\ ACS Ultra Deep Field \citep{Beckwith06-ACSUDF} 
in the filters F606W and F775W.

\begin{figure}
\begin{picture}(17.4,9.4)
\put(0.1,0.){{\psfig{figure=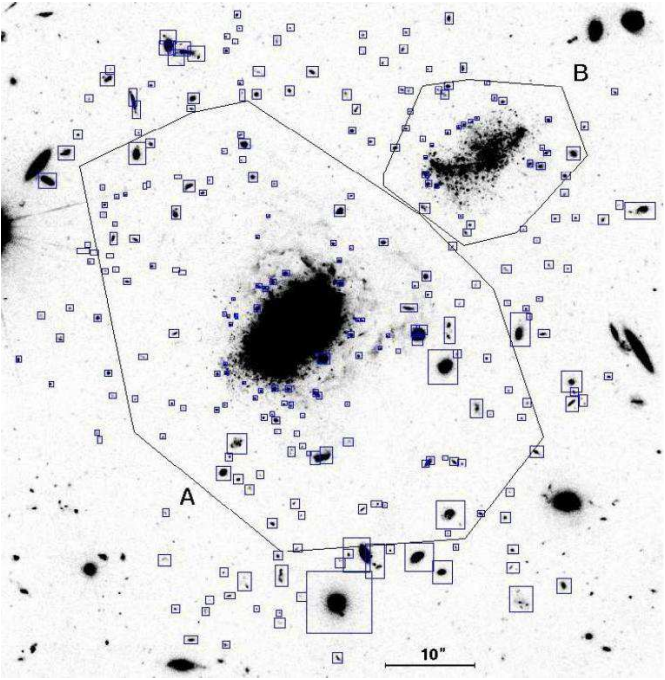,width=8.8cm,angle=0,clip=}}}
\end{picture}
\caption[]{Combined $I$ band exposure of \iz18\ and \cc\ with point- and compact sources 
in the vicinity of the galaxies (polygonal regions A and B) marked with
rectangles. North is up and east to the left.  
}
\label{fig:IZw18_cps}
\end{figure}

The data processing was carried out using 
IRAF\footnote{IRAF is the Image Reduction and Analysis Facility distributed by
the National Optical Astronomy Observatory, which is operated by the
Association of Universities for Research in Astronomy (AURA) under
cooperative agreement with the National Science Foundation (NSF).} and ESO
MIDAS\footnote{Munich Image Data Analysis System, provided by the European
Southern Observatory (ESO)}. Photometric quantities refer to the Vega system. 

Since the main goal of this study is deep surface photometry, its most
critical aspect is the removal of diffuse and compact background
sources (diffuse extragalactic background, zodiacal light, foreground
stars and background galaxies, respectively). 
After subtraction of the diffuse background, we therefore checked to 
which extent compact or point sources (\cps) within the extended 
LSB envelope of \iz18\ can affect SBPs and color profiles. 
For example, for an extended source of constant surface brightness 
$\mu=29$ \sbb, the total apparent magnitude $m$ of a circular annulus with 
19\arcsec$\leq$\rr$\leq$20\arcsec\ (roughly the radius of \iz18) is 23.8 mag. 
At this intensity level, already a single faint ($m$=25 mag) background
\cps\ can introduce an error of 0.3 mag in surface photometry.
We adopted the following procedure: after image alignment and correction
for cosmics, we used the combined images in the three filters to compile 
a catalogue of \cps\ in the relevant portion of the field of view.
In doing this, we disregarded \cps\ in \iz18\ (roughly the area subtended by 
the 25 $R$\arcmin\ \sbb\ isophote in Fig. \ref{IZw18HaImage}) and in \cc, 
as well as compact clumps of \ige, identified by their 
blue $V$--$I$ and $R$--$I$ color.
All \cps\ in each frame were in turn replaced by the mean intensity  
in the adjacent area. 
Special care was given to the removal of two background galaxies
close to the western super-shell of \iz18. This was done by subtracting 
a 2D model, computed with the \cite{BenderMollenhoff} algorithm. 
In total, $\sim$140 \cps\ were removed in the field of interest around 
\iz18\ and \cc\ (polygonal regions labeled A and B in Fig. \ref{fig:IZw18_cps}).
Their integral $I$ magnitude of 19.17 mag and 21.3 mag within 
the regions considered ($\sim$1600 $\sq\arcsec$ and $\sim$200
$\sq\arcsec$ for A and B, respectively) corresponds to a mean 
surface brightness of $\sim$27 $I$ \sbb.  
This value is consistent at the 1.5$\sigma$ level with the value of 26.7 $I$ \sbb\ 
inferred by \citet{ZMO09} for the resolved extragalactic background light emission.
The isophotal radius of the \emph{stellar} component of \iz18\ and \cc\ at 
$\mu$=26 $B$ \sbb\ was determined by P02 to be 8\farcs8 and 5\farcs5,
respectively. From their combined isophotal area of $\sim$340 $\sq\arcsec$ 
and the above derived \cps\ surface density of $\sim$0.08 $\sq\arcsec^{-1}$, we 
expect about 27 background \cps\ in \iz18\ and \cc\ altogether.
SBPs computed prior to and after removal of \cps, were found in all bands to be 
consistent within 1$\sigma$ uncertainties, ensuring that \cps\ contamination 
does not affect our conclusions in Sect. \ref{results}. 

SBPs were computed with the code {\nlx iv} (P02) \citep[also referred to as
{\tt Lazy} by][]{Noeske06-UDF} that was specifically developed for the study 
of irregular galaxies.
This code permits a simultaneous processing of co-aligned images of a galaxy 
in several bands and does not require a choice of a 
galaxy center, neither does implicitly assume that the galaxy can be approximated
by the superposition of axis-symmetric luminosity components.
One of its key features is the computation of photon statistics within automatically 
generated irregular annuli that are adjusted to the galaxy morphology for each 
surface brightness interval $\mu\pm\Delta\mu$. 
This distinguishes code {\nlx iv} from other surface photometry packages
which generally employ ellipse-fitting to isophotes or photon
statistics within elliptical annuli (e.g. {\nlx meth. i} of P96a, task ELLIPSE in
IRAF, FIT/ELL3 in MIDAS), or approximate a galaxy by a single or several 2D
axis-symmetric components (e.g. GIM2D, \citet{Simard98} and GALFIT, \citet{Peng02}).

\section{Results \label{results} }

\subsection{The photometric structure of I\ Zw\ 18\,C \label{phot:IZw18C} }

\begin{figure*}
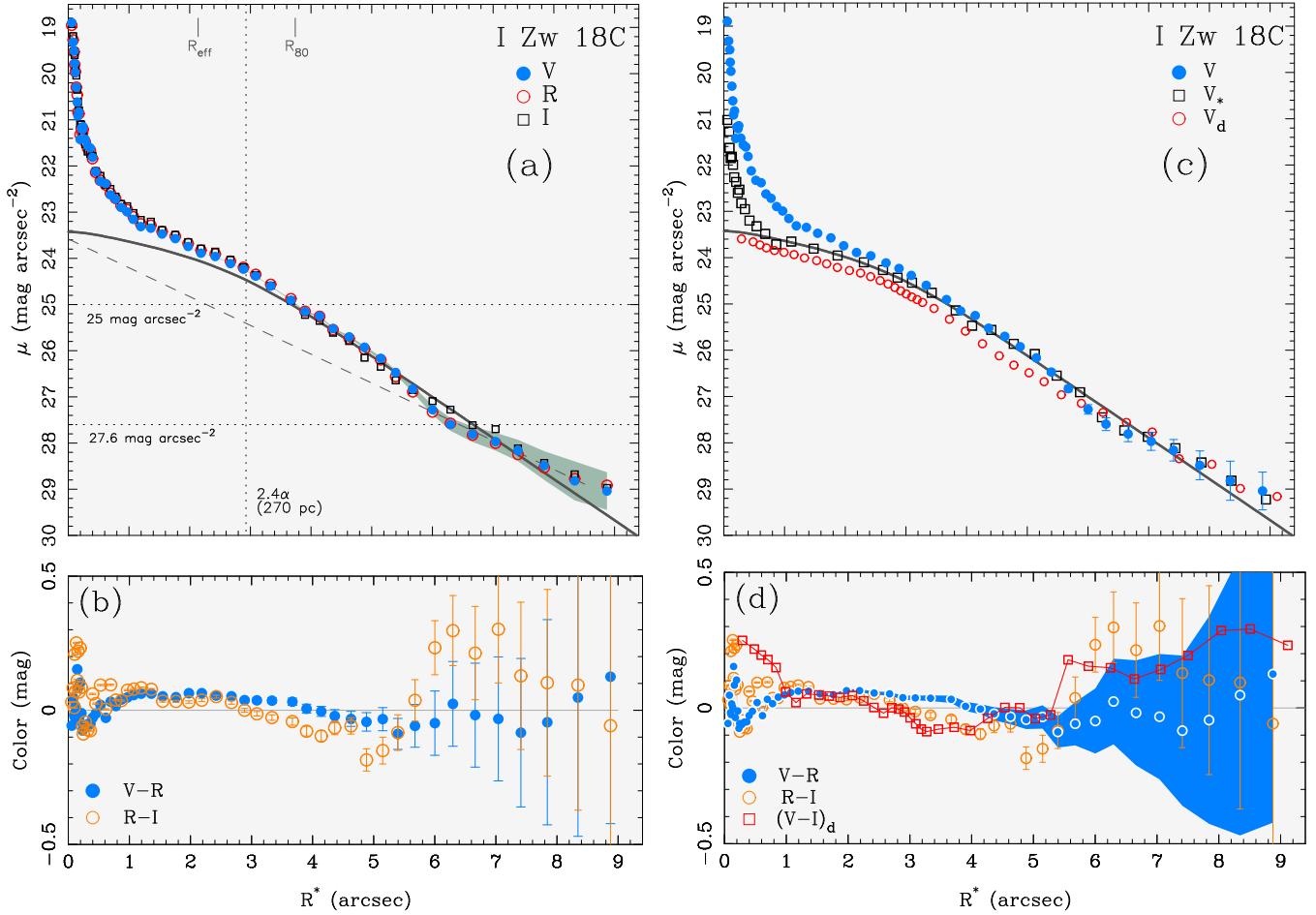

\begin{picture}(16.4,12.6)
\put(0,5.0){{\psfig{figure=fig3a.ps,width=8.7cm,angle=-90}}}
\put(0.05,0.){{\psfig{figure=fig3b.ps,width=8.7cm,angle=-90}}}
\put(9,5.0){{\psfig{figure=fig3c.ps,width=8.7cm,angle=-90}}}
\put(9.05,0.){{\psfig{figure=fig3d.ps,width=8.8cm,angle=-90}}}
\end{picture}
\caption[]{{\bf (a)} Surface brightness profiles (SBPs) of \cc\ in $V$
(F555W), $R$ (F606W) and $I$ (F814W). The thick gray curve shows a fit 
to the $V$ SBP for \rr$\geq$3\farcs8 with the modified exponential fitting 
function Eq. \ref{eq:p96a} ({\tt modexp}) for a core radius 
$R_{\rm c}=2.4\cdot\alpha$ (dotted vertical line). 
The effective radius $R_{\rm eff}$ and the radius $R_{80}$ enclosing
80\% of the total $V$ emission are indicated.
The dashed line shows a linear fit to the outermost exponential part
of the $V$ SBP for \rr$\geq$6\arcsec, i.e. at the extremely faint 
($\mu\geq 27.6$ \sbb) outskirts of the galaxy.
The shadowed area corresponds to the 1$\sigma$ uncertainties of the $V$ SBP.
{\bf b)} Radial $V$--$R$ and $R$--$I$ color profiles of \cc, derived from the 
SBPs in the upper panel.
{\bf c)} Comparison of the best-fitting {\tt modexp} model in panel a (thick
gray curve) with the $V$ SBPs of \cc\ after partial removal of the brightest 
point sources ($V_{\star}$; squares). 
Open circles show the SBP of the \emph{unresolved} stellar emission
($V_{\rm d}$), computed after complete removal of compact ($\leq$0\farcs5)
sources with an unsharp masking technique.
{\bf d)} Comparison of the color profiles in panel b (open and filled circles) 
with the ($V$--$I$)$_{\rm d}$ color profile (squares) of the unresolved
stellar emission. 
The shadowed area depicts the 1$\sigma$ uncertainties of the $V$--$R$ profile.}
\label{IZw18C_sbp}
\end{figure*}

%
\begin{figure*}
\begin{picture}(16.4,17.7)
\put(0,0){{\psfig{figure=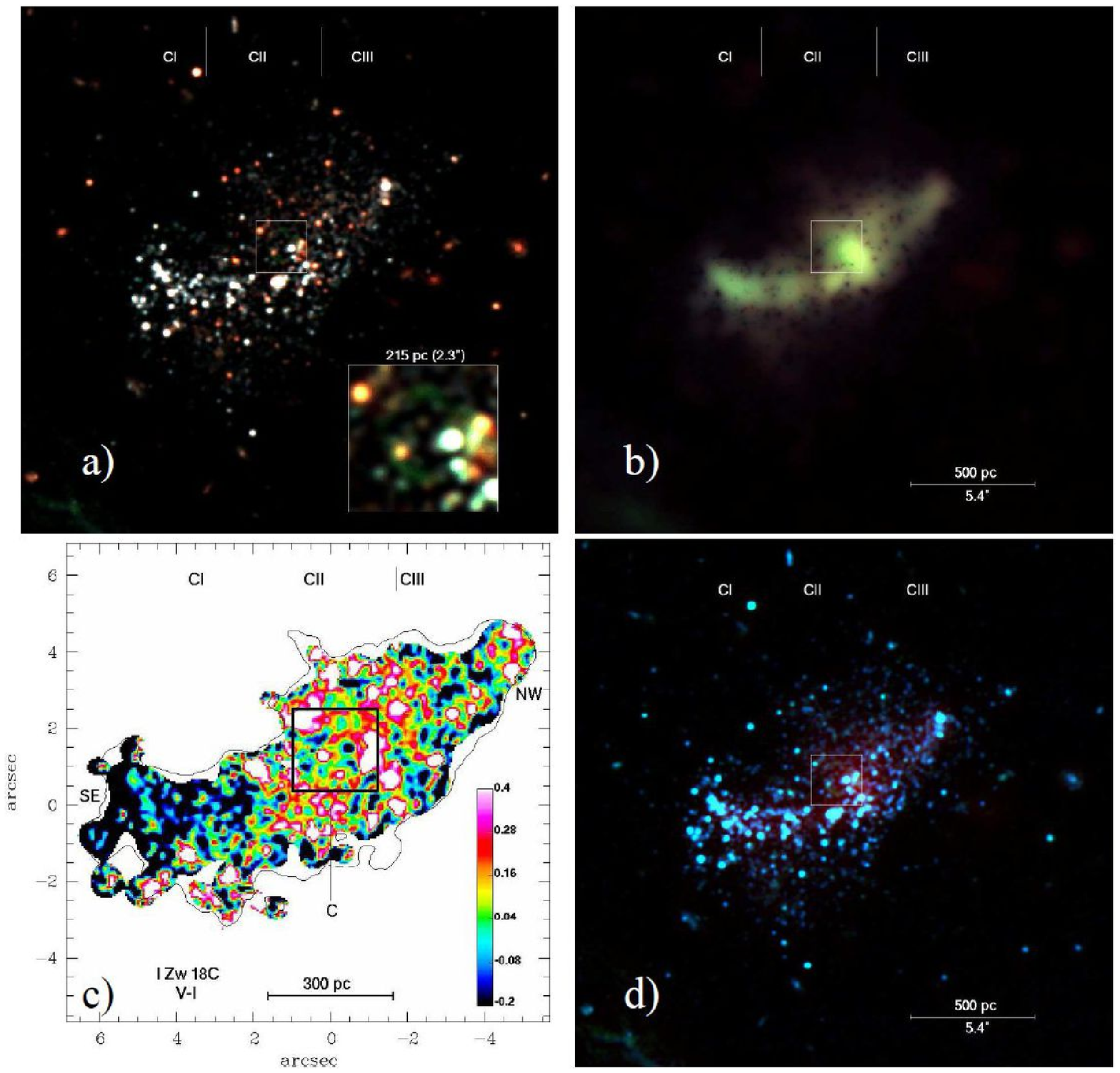,width=18.4cm,angle=0,clip=}}}
\end{picture}
\caption[]{{\bf (a)} Three-color composite of the images $R_{\rm c}$, $V_{\rm c}$ and 
$I_{\rm c}$ (red, green and blue image channel, respectively), illustrating the
  spatial distribution of compact ($\leq$0\farcs5) sources in \cc. 
The regions C\,{\sc i}, C\,{\sc ii} and C\,{\sc iii} defined by \cite{IT04-IZw18} are indicated.
The blown up version of the central part of \cc\ (lower-right) shows the
ionized gas shell in the vicinity of the bright stellar cluster C
\citep[cf][]{Dufour96b}.
{\bf (b)} Three-color composite of $R_{\rm d}$, $V_{\rm d}$ and $I_{\rm d}$,
of the \emph{unresolved} stellar emission. The color coding is the same as in panel a.
{\bf c)} $V$--$I$ color map of \cc, revealing very blue colors ($V$--$I<$0)
all over the southeastern third (region C\,{\sc i}) of the galaxy.
The contour corresponds to 25 $V$ \sbb.
{\bf d)} Comparison of the spatial distribution of compact and point sources in
$V_{\rm c}$ and $R_{\rm c}$ (blue and green channel, respectively) 
with the unresolved stellar background ($I_{\rm d}$, red channel).
}
\label{IZw18C_hb}
\end{figure*}

In this section we investigate the photometric structure of \cc, based on the combined 
\hst\ ACS data in the filters $V$, $R$ and $I$. These allow to extend previous \hst\ WFPC2
surface photometry (P02) by $\ga$1 mag, with significantly reduced uncertainties below $\mu\sim26$ \sbb. 
In agreement with previous work, the SBPs of \cc\ in all bands (Fig. \ref{IZw18C_sbp}a) 
were found to be nearly indistinguishable from one another, implying nearly constant
radial colors. All SBPs display a narrow (\rr$\leq$1\arcsec) central excess and an 
outer (3\arcsec$\la$\rr$\la$9\arcsec) roughly exponential drop-off that mainly
reflects the luminosity output from the host galaxy.
A salient feature at intermediate radii (1\arcsec$\leq$\rr$\la$3\arcsec) 
is a shallower intensity increase than what inwards extrapolation of the outer 
profile slope to \rr=0\arcsec\ predicts.
An adequate fit to such SBPs therefore requires a modified
exponential model that involves an extended (\rr$\sim$3\arcsec) central core 
of nearly constant surface brightness.
One such fitting function (hereafter {\tt modexp}), used by P96a 
to fit the host galaxy of BCDs 
\citep[see][for applications on near infrared (NIR) studies and a comparison 
with the S\'ersic law]{Noeske03-NIR} has the form:

\begin{equation}
I(R^*) = I_{\rm exp} \cdot
\big[1-\epsilon_1\,\exp(-P_3(R^*))\big],
\label{eq:p96a} 
\end{equation}
where $P_3(R^*)$ is defined as
\begin{equation}
P_3(R^*) =
\left(\frac{R^*}{\epsilon_2\,\alpha}\right)^3+\left(\frac{R^*}{\alpha}\,\frac{1-\epsilon_1}{\epsilon_1}\right).
\label{eq:p96b} 
\end{equation}

In addition to the central intensity $I_0$ and scale length $\alpha$ of a pure
exponential profile $I_{\rm exp}=I_0\cdot\exp(R^*/\alpha)$, Eqs. \ref{eq:p96a}\&\ref{eq:p96b} 
involve two further parameters: the central intensity depression 
$\epsilon_1=\Delta I/I_0$ relative to an exponential model and the 
core radius $R_{\rm c}=\epsilon_2 \cdot \alpha$.
The best-fitting {\tt modexp} model to the $V$ band SBP for $\epsilon_{1,2}$=(2.4,0.8) (cf P02) 
is shown in Fig. \ref{IZw18C_sbp}a with the gray thick curve.
It yields in all bands an \emph{extrapolated} central surface brightness 
$\mu_{\rm E,0}$=21.7$\pm$0.2, a \emph{true} central surface brightness 
$\mu_0=\mu_{\rm E,0}-2.5\,log(1-\epsilon_2)$=23.45 \sbb\ and an $\alpha$ in
the narrow range between 108 and 117 pc. The absolute $V$ magnitude
determined from the {\tt modexp} fit (--11.67 mag) corresponds to
$\sim$80\% of the total luminosity of \cc.

Note that a direct determination of the intensity profile of the host galaxy
of BCDs for radii \rr$\leq R_{\rm c}$ is generally prevented by the luminous
young stellar component that typically dominates out to 
\rsf$\approx$2$\alpha$ \citep[][hereafter P96b]{P96b}.  
Since young stellar clusters (SCs) can hardly be sufficiently resolved and 
subtracted out, even at the angular resolution of the \hst, the chosen parameter set
$\epsilon_{1,2}$ has to rely on plausibility arguments 
\citep[see discussion in P96a and][]{Noeske03-NIR} 
and is to be considered approximative only.

In the case of \cc, however, due to the comparatively low surface density of
young SCs, one stands a better chance to directly constrain the central form
of the host's SBP.
For this, we first subtracted out the brightest $\sim$200 point sources from
the galaxy using DAOPHOT \citep{Stetson79-DAOPHOT} and subsequently recomputed 
the $V$ SBP from the residual emission. 
Figure \ref{IZw18C_sbp}c shows that this SBP (open squares, labeled $V_{\star}$) 
closely matches the best-fitting {\tt modexp} model for intermediate to large \rr.
This agreement suggests that the adopted $\epsilon_{1,2}$ parameters yield a
reasonable first-order approximation to the unresolved emission of the host.
SBP integration indicates that roughly 30\% of \cc's emission within 
$R_{80}$ (Fig. \ref{IZw18C_sbp}) is due to point sources.
This is rather a lower limit for the luminosity fraction of \cps\
given that with the adopted procedure SCs could not be fully deblended and
subtracted out. This is apparent from the still strong ($\sim$2 mag) central peak 
of the $V_{\star}$ SBP that is mainly due to incomplete removal of the central 
SC complex C and surrounding bright SCs (cf Fig. \ref{IZw18C_hb}a).

In an effort to better constrain the SBP of the \emph{unresolved} stellar
component of \cc, we subsequently applied a flux-conserving unsharp masking technique 
\citep[][hereafter P98]{Papaderos98-SBS0335} to filter out all
higher-surface brightness (HSB) clumpy features with $\leq$0\farcs5 and
isolate the diffuse $V$ emission only.
The frames holding the compact ($V_{\rm c}$, $R_{\rm c}$ and $I_{\rm c}$)
and diffuse ($V_{\rm d}$, $R_{\rm d}$ and $I_{\rm d}$) emission in each band
are displayed in panels {\nlx a} and {\nlx b} of Fig. \ref{IZw18C_hb}, respectively.
The $V_{\rm d}$ SBP (open circles in Fig. \ref{IZw18C_sbp}c) 
is at intermediate radii (1\arcsec$\la$\rr$\la$6\arcsec) fairly comparable to the 
$V_{\star}$ SBP, except for a nearly constant offset by $\approx$0.3 mag. 
Its corresponding absolute magnitude (--11.3 mag) is by about 0.75 mag fainter 
than the integral value for \cc\ (--12.05 mag) and can be regarded as 
characteristic for the \emph{unresolved} stellar emission in the host galaxy.

Note that all SBPs in Figs. \ref{IZw18C_sbp}a,c display at very faint levels 
($\mu\ga 27.6$ \sbb) a shallower outer (\rr$\ga$6\arcsec) 
exponential slope (dashed line in Fig. \ref{IZw18C_sbp}a). 
This feature is certainly not due to point spread function (PSF) convolution 
effects since the maximum extent of the ACS PSF at its lowest 
measured intensity (10 mag below its central value) is $\sim$1\farcs5
\citep{Jee07-ACSPSF}. \ige\ contamination from the main body
can as well be excluded both on the basis of narrow-band imaging
\citep[e.g.][P02]{Ostlin96} and because it would bluen $V$--$I$ and
$R$--$I$ color profiles, in disagreement with the slight reddening
of color profiles for \rr$\ga$6\arcsec\ (see below).  
This outermost SBP feature has evaded detection on previous surface 
photometry (P02) which, due to photometric uncertainties,
was practically restricted to within the Holmberg radius.
Because of its faintness (merely 3\% of the total emission) and
low surface brightness ($\sim$27.6 -- 29 \sbb), its reality can 
not be established beyond doubt even with the present data. 
If due to an underlying stellar host of constant exponential
slope to \rr=0\arcsec, its $\mu_0$ ($\approx 23.6$ \sbb) and $\alpha$ (160 pc) 
would qualify it as an LSB dwarf with a $M_V\approx$--11 mag 
(equivalent to $\sim$38\% of \cc's emission). 

We turn next to the color distribution in \cc. 
Figure \ref{IZw18C_sbp}b shows that for 1\arcsec$\la$\rr$\la$6\arcsec\ both 
the $V$--$R$ and $R$--$I$ index is very blue and nearly constant 
($\approx$0 with a standard deviation about the mean of 0.05 mag). 
Since \rr$=$6\arcsec\ encompasses practically the total emission of \cc, 
this color may be regarded as representative for the galaxy as a whole. 
In the outermost periphery of the galaxy, where the shallower outer exponential
component appears, color profiles hold a hint for a slightly redder 
$R$--$I$ (0.16$\pm$0.09) without a notable change in $V$--$R$.
The color profiles in their innermost (\rr$\leq$1\arcsec) part are
dominated by luminous \cps\ in the surroundings of region C (cf Fig. 1) 
and show a large scatter (up to 0.15 mag) around mean values of 0 and 0.06 mag
for $V$--$R$ and $R$--$I$, respectively.

In order to place meaningful constraints on the evolutionary status of \cc,
it is necessary to verify that these blue colors are not due to the 
luminosity-weighted average of bright blue SCs with a red underlying stellar
background. The latter could readily escape detection, even when dominating 
the stellar mass (see e.g. P98).
This concern is underscored by two atypical properties of \cc:
First, the galaxy exhibits a weak color gradient along its major 
axis \cite[cf][]{Aloisi99-IZw18,Izotov01-IZw18}, with its northwestern tip 
being redder ($B-V$ = 0.05 mag, $V-I$ = 0.2 mag) than the southeastern one 
($B-V$ = --0.07 mag and $V-I$ = --0.2 mag). 
This is apparent from Fig. \ref{IZw18C_hb}c, from which a $V$--$I$ color as blue as
--0.2 mag can be read off all over the southeastern third of the galaxy 
(region C\,{\sc i} in the denomination by IT04).
By contrast, the average $V$--$I$ color in the central (C\,{\sc ii}) and 
northwestern (C\,{\sc iii}) part of the galaxy is redder ($\geq$0 mag) 
with several local color maxima ($\geq$0.3 mag) associated with \cps.
It is thus conceivable that the bluer southeastern and redder northwestern galaxy half
counter-balance each other, thereby introducing an overall blue mean radial color.
Another characteristic of \cc\ that could additionally conspire in diminishing 
radial color gradients is that the surface density of its \cps\ tends to be
spatially anti-correlated with the \emph{unresolved} stellar background of the host.
This is illustrated in Fig. \ref{IZw18C_hb}d where the diffuse emission of the
latter ($I_{\rm d}$: red channel) is overlaid with the \cps\ in the  
$V_{\rm c}$ and $R_{\rm c}$ images (blue and green channel, respectively): 
it can be seen that the diffuse component peaks at the northwestern half of \cc\
(C\,{\sc ii} and C\,{\sc iii}) in which it accounts for $\approx$50\% 
of the $V$ line-of-sight intensity, whereas its contribution drops to 
$\la$30\% in the southeastern half of \cc\ (region C\,{\sc i}) where most of the
bright blue \cps\ are located.
Consequently, especially in region C\,{\sc i}, a hypothetical red stellar
background could readily escape detection on radial color profiles.

From such considerations, and in order to infer the colors of the
unresolved stellar component of \cc\ in an as unbiased manner as possible, we additionally 
computed color profiles based on $V_{\rm d}$, $R_{\rm d}$ and $I_{\rm d}$ SBPs only.
In all cases, we found a good agreement with the results initially obtained from 
the total emission with the exception of a central (\rr$\la$1\arcsec) red peak 
with mean values of ($V$--$R$)$_{\rm d}$=0.1$\pm$0.03 and 
($V$--$I$)$_{\rm d}$=0.16$\pm$0.05.
This color excess might be attributed to a stellar age gradient or 
enhanced extinction in region C and
surroundings. Note that \cite{Izotov01-IZw18} estimated from spectral synthesis
models an extinction coefficient C(H$\beta$)=0.1--0.3 for region C,
corresponding to $A_V$=0.2--0.65 mag.
%
\begin{figure}
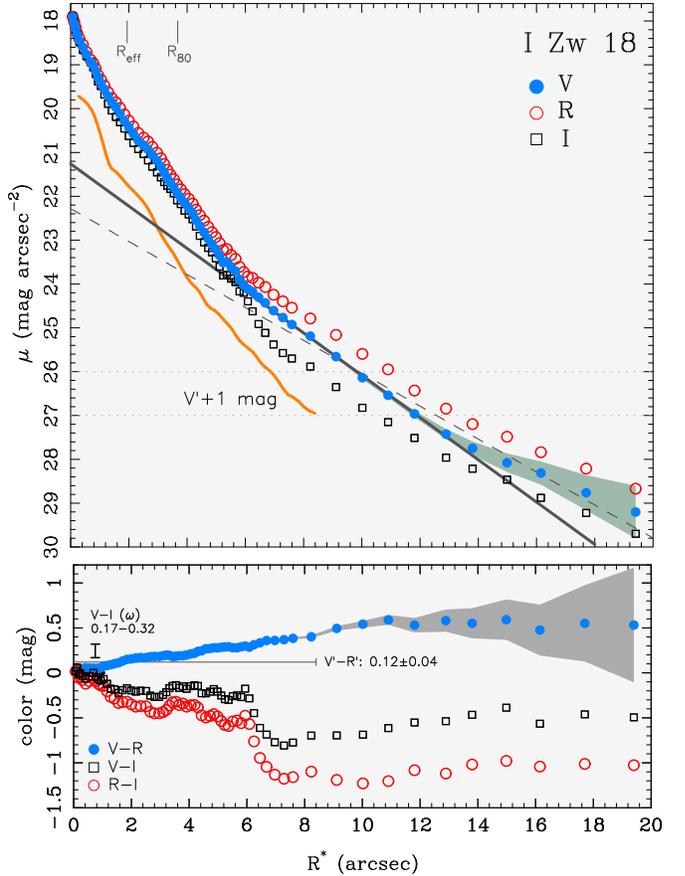

\begin{picture}(8.6,11.8)
\put(0,4.2){{\psfig{figure=fig5a.ps,width=8.6cm,angle=-90}}}
\put(0.1,0){{\psfig{figure=fig5b.ps,width=8.6cm,angle=-90}}}
\end{picture}
\caption[]{
{\bf (upper panel)} $V$, $R$ and $I$ SBPs of \iz18. The thick-solid and
thin-dashed lines show linear fits to the $V$ SBP in the radius range
7\arcsec$\leq$\rr$\leq$15\arcsec\ and \rr$\geq$7\arcsec, respectively.
The effective radius R$_{\rm eff}$ and the radius R$_{\rm 80}$ enclosing 80\% 
of the total $V$ emission are indicated.
The $V$ SBP derived by P02 after subtraction of the \hb\ and 
[O{\sc iii}]$\lambda\lambda$4959,5007 emission lines (referred to as
$V$\arcmin), shifted by +1 mag for the sake of clarity, is included for comparison.
{\bf (lower panel)} $V$--$R$ (filled circles), $V$--$I$ (squares) and $R$--$I$
(open circles) color profiles, computed from the SBPs in the upper panel.
The vertical bar indicates the $V$-$I$ color range inferred by P02 from ground 
based and \hst\ PC data for region $\omega$ at the southeastern tip of \iz18\
($V$-$I$($\omega$) = 0.17 \dots 0.32 mag). 
Since nebular emission is relatively weak in that region, its color can be considered 
representative for the stellar host of \iz18.
The mean $V$--$R$ color of \iz18's host after subtraction of nebular line emission 
($V$\arcmin--$R$\arcmin=0.12$\pm$0.04 mag; P02) is indicated by the horizontal
line. Note that the colors $V$-$I$($\omega$) and $V$\arcmin--$R$\arcmin\ 
of the stellar host galaxy are by, respectively, $\ga$0.8 mag redder and
$\approx$0.4 mag bluer than the colors of the exponential nebular envelope 
(\rr$\geq$6\arcsec).}
\label{IZw18SBP}
\end{figure}

As apparent from panel d of Fig. \ref{IZw18C_sbp}, the $V$--$I$ profile of
the unresolved stellar component (($V$--$I$)$_{\rm d}$, squares) is fairly comparable
to the $V$--$R$ and $R$--$I$ profiles (panel b), revealing a nearly constant color of 
$\approx$0$\pm$0.04 mag within 1\arcsec$\leq$\rr$\leq$6\arcsec\
and a slightly redder value (0.2$\pm$0.08 mag) in the extreme periphery of the
galaxy.
We are therefore led to conclude that the overall blue colors of \cc\ 
for 1\arcsec$\la$\rr$\la$6\arcsec are not dictated by bright young 
SCs spread all over the body of the galaxy but as well characteristic for its 
unresolved stellar component.
  
\subsection{The photometric structure of I\ Zw\ 18 \label{phot:IZw18} }
%
\begin{figure*}[ht]
\begin{picture}(16.4,9.7)
\put(0,0){{\psfig{figure=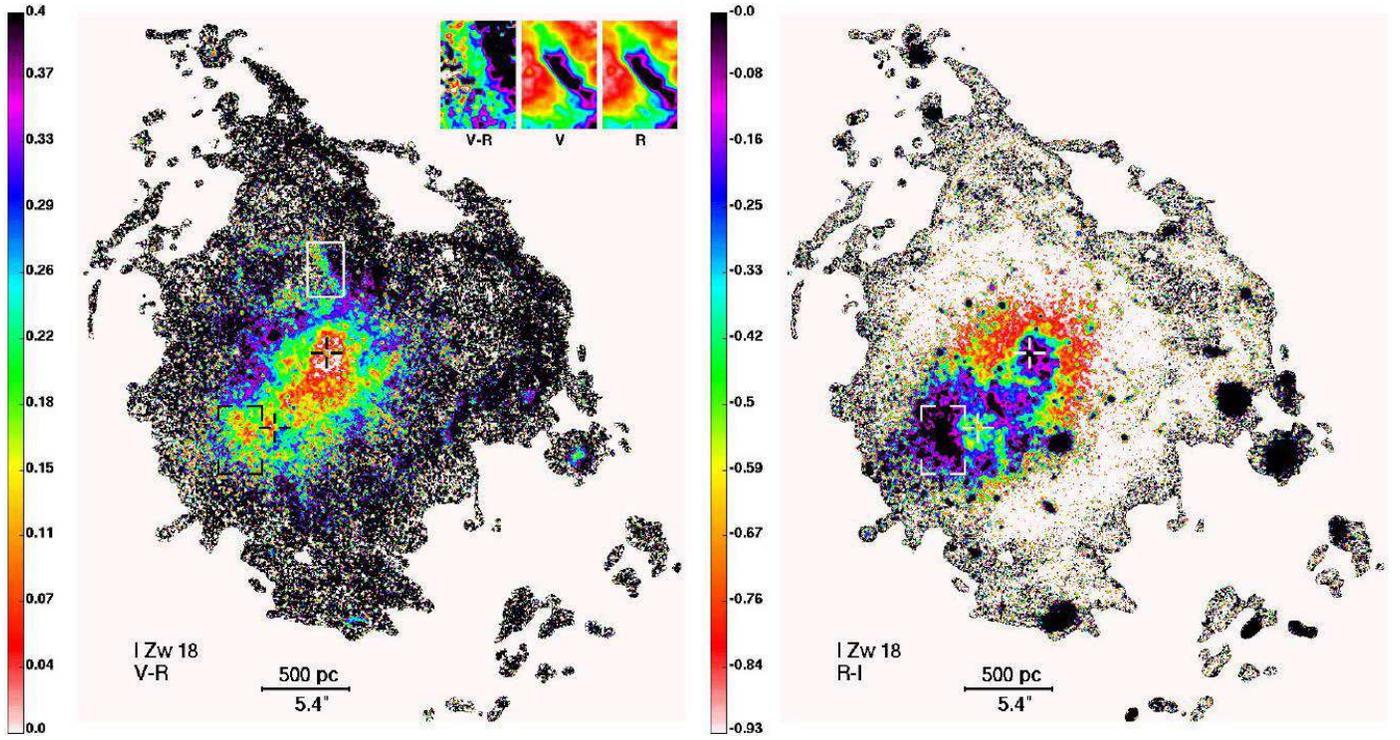,width=18.4cm,angle=0}}}
\end{picture}
\caption[]{$V$--$R$ (left panel) and $R$--$I$ (right panel) color map of 
\iz18, displayed in the range between 0 and 0.4 mag and --0.9 and 0 mag,
respectively. Crosses mark the star-forming regions NW and SE. 
The region termed $\omega$ by P02 (cf Fig. \ref{IZw18-Fig1}) 
is indicated at the southeastern part of the image. 
The insets in the left panel show a magnified view of a shell 
$\approx$5\arcsec\ northwards of region NW.
}
\label{IZw18-VRImaps}.
\end{figure*}
The $V$, $R$ and $I$ SBPs of \iz18\ (Fig. \ref{IZw18SBP}) compare well to those
presented by \ko\ and P02. They exhibit a central (\rr$\la$6\arcsec) 
high-surface brightness core and a nearly exponential LSB envelope extending out
to \rr$\sim$20\arcsec. 
Profile fitting in the radius range 7\arcsec$\leq$\rr$\leq$15\arcsec\ (solid line)
yields for the LSB component an $\alpha=2\farcs3$ (210 pc) and a luminosity fraction 
of $\approx$30\%.
At fainter levels ($\ga$ 27.6 -- 30 \sbb) all SBPs exhibit a shallower 
exponential slope with $\alpha=4\farcs1$ (370 pc).
This SBP feature is attributable to the faint filamentary emission in the extreme 
northwestern and southeastern 
periphery of \iz18\ (cf. Fig. \ref{IZw18HaImage}) that has evaded 
detection on previous \hst\ WFPC2 imagery and which accounts for no more than 
1\% of the total luminosity. A linear fit to the whole LSB envelope
(7\arcsec$\leq$\rr$\leq$20\arcsec; dashed line) yields a mean 
$\alpha=2\farcs9\pm 0\farcs13$ ($\sim$270 pc). 

The radial color profiles of \iz18\ (lower panel of Fig. \ref{IZw18SBP}),  
reflect the increasing contribution of \ige\ to the observed
line-of-sight intensity with increasing galactocentric radius, 
in agreement with previous evidence (P02).
The $V$--$R$ color (filled circles) increases roughly
linearly from $\la$0 at \rr=0\arcsec\ to 0.55 mag at \rr=9\arcsec\ 
where it levels off to a nearly constant value. 
The corresponding, relatively strong color gradient of 
$\gamma_+$=0.6 mag kpc$^{-1}$  is rather typical for evolved BCDs (P96b), 
suggesting, at first glance, that SF activity in \iz18\ occurs within 
a more extended, old underlying host.
This interpretation is, however, immediately challenged upon inspection 
of the extremely blue $V$--$I$ and $R$--$I$ profiles.
The latter display already in their inner part (2\arcsec$\la$\rr$\leq$6\arcsec) 
mean values of --0.21 mag and --0.44 mag, respectively, both inconsistent 
with the red $V$--$R$ color, if the emission is assumed to be dominated by stars.
More impressively, at \rr$\sim$6\arcsec, i.e. roughly at the transition 
radius between the HSB core and the exponential LSB envelope of SBPs, either
color index shows a sudden decrease to values as blue as 
$V$--$I$=--0.61$\pm$0.13 mag and $R$--$I$=--1.1$\pm$0.08 mag, 
which then remain nearly constant out to \rr$\sim$20\arcsec.
\begin{figure}
\begin{picture}(8.9,8.0)
\put(0,0){{\psfig{figure=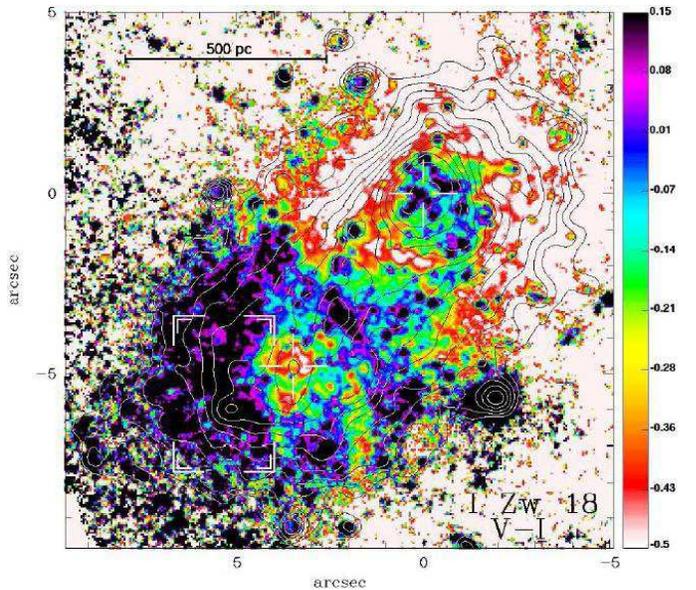,width=8.9cm,angle=0,clip=}}}
\end{picture}
\caption[]{$V$--$I$ map of the central 15\farcs6$\times$15\farcs6
of \iz18, displayed in the color range between --0.5 and 0.15 mag. The
contours are computed from the $R$\arcmin\ image by P02 (see also
Fig. \ref{IZw18HaImage}) and go from 19 to 25 \sbb\ in steps of 0.5 mag.
Note the very blue ($<$--0.5 mag) rim encompassing the northwestern 
star-forming region.
The mean $V$--$I$ color over the southeastern region $\omega$ is determined to be 
0.23 mag with an upper value of 0.3 mag in its reddest quarter.}
\label{IZw18-VImap}
\end{figure}

That such colors can not be of stellar origin is apparent already from the
fact that, even for an O5V star, the $V$--$I$ and $R$--$I$ color is --0.32 mag and
--0.18 mag, respectively, i.e. 0.3 to 0.9 mag redder than the color of the
LSB envelope of \iz18. More generally, as pointed out by P02, there 
is no stellar population, regardless of star formation history, age and 
metallicity that can reproduce the observed combination of red ($\sim$0.6 mag) 
$V$--$R$ with blue (--0.6 \dots --1.1 mag) $V$--$I$ and $R$--$I$ 
colors of the LSB envelope of \iz18. 
Quite to the contrary, such colors can solely be accounted for by \ige.
\begin{figure*}
\begin{picture}(17.4,16.8)
\put(0.2,0){{\psfig{figure=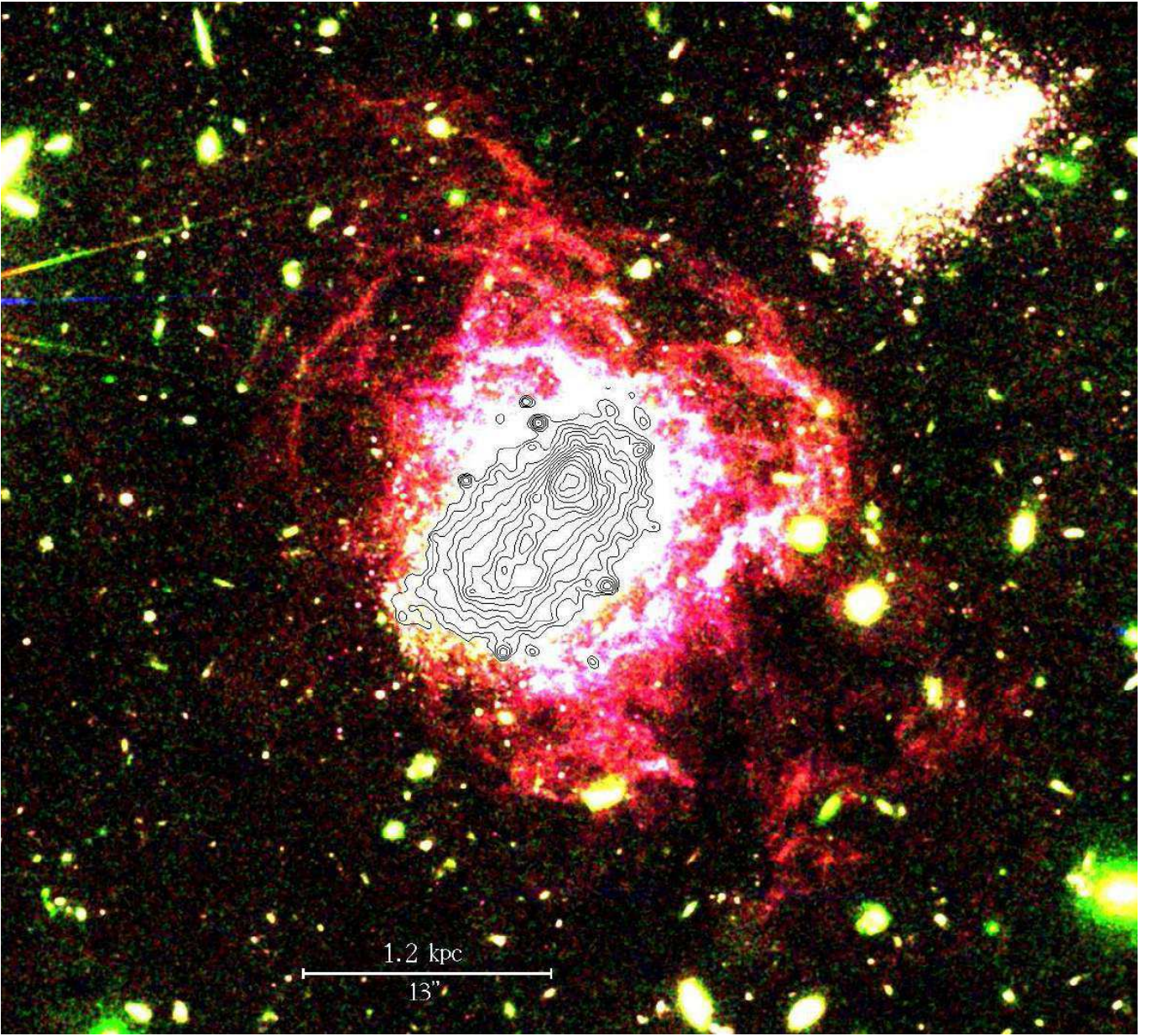,width=18.0cm,angle=0,clip=}}}
\end{picture}
\caption[]{
Composite image of \iz18\ and \cc\ with the $R$, $I$ and $V$ bands shown
in the red, green and blue channel, respectively.
The image discloses a complex network of ionized gas filaments extending as far
out as 2.6 kpc, approximately twice the distance reported from previous 
\hst\ WFPC2 studies, corresponding to  
$\sim$16 exponential scale lengths $\alpha$ of the \emph{stellar} host
galaxy of \iz18\ (160 pc; P02). 
The contours, adapted from P02, are computed from \hst\ WFPC2 $R$ data, 
after two-dimensional subtraction of the \ha\ emission 
(referred to as $R$\arcmin), they thus delineate the morphology of the 
stellar and nebular continuum emission in the main body of \iz18. 
Contours go from 19 to 25 $R$\arcmin\ in steps of 0.5 mag.
Note that \iz18\ (i.e. its \emph{stellar} component, as depicted by the 
$R$\arcmin\ contours) and \cc\ show a remarkably similar structure.
}
\label{IZw18HaImage}
\end{figure*}

Indeed, comparison of Fig. \ref{IZw18SBP} with Fig. 12 of P02 reveals that 
the radius \rr=6\arcsec -- 7\arcsec\ where the steep  $V$--$I$ and $R$--$I$ color drop-off occurs
is identifiable with the radius where the line-of-sight contribution of stellar emission 
steeply decreases from a plateau value of $\sim$40\% for
3$\la$\rr$\la$6\arcsec\ to less than 20\%.
Therefore, the much deeper data studied here confirm and strengthen the previous
conclusion that the extended (6\arcsec$\la$\rr$\la$20\arcsec) exponential LSB envelope of
\iz18\ is due to \ige.

The gaseous nature of the LSB envelope of \iz18\ is also evident from the 
color maps in Figs. \ref{IZw18-VRImaps} and \ref{IZw18-VImap}.
These show a clear correspondence to radial color profiles, most notably 
a strong core--envelope color contrast (0.5 -- 1.5 mag) and remarkably 
uniform colors for the envelope
($V$--$R$$\sim$ 0.5 \dots 0.6, $R$--$I$$\approx$--1.4 mag, $V$--$I$$\approx$--1
mag) over a spatial scale of 9 kpc$^2$.
The northwestern super-shell, for example, though prominent on direct images (cf Fig. 1),
is barely distinguishable from its gaseous surroundings on 
$V$--$I$ and $R$--$I$ color maps, suggesting a nearly constant spectral energy
distribution (SED) all over the LSB envelope.

The relative extent of the stellar component that is confined to the compact
HSB core with respect to the nebular LSB envelope is better 
illustrated in Fig. \ref{IZw18HaImage}. 
The image reveals a complex network of ionized gas filaments extending as far
out as 2.6 kpc, twice the galactocentric distance previously 
reported from \hst\ WFPC2 studies and equivalent to $\sim$16 exponential 
scale lengths $\alpha$ of the \emph{stellar} host of \iz18\ (160 pc; P02). 
The contours, adapted from P02, are computed from \hst\ WFPC2 $R$ data, 
after two-dimensional subtraction of \ha\ line emission 
(referred to as $R$\arcmin), they thus delineate the morphology of the 
stellar and nebular continuum emission.
Note that \iz18\ (i.e. its \emph{stellar} component as depicted by the
$R$\arcmin\ contours) and \cc\ show a remarkably similar structure.
\begin{figure*}[!ht]
\begin{picture}(16.4,16.8)
\put(0.2,0.0){{\psfig{figure=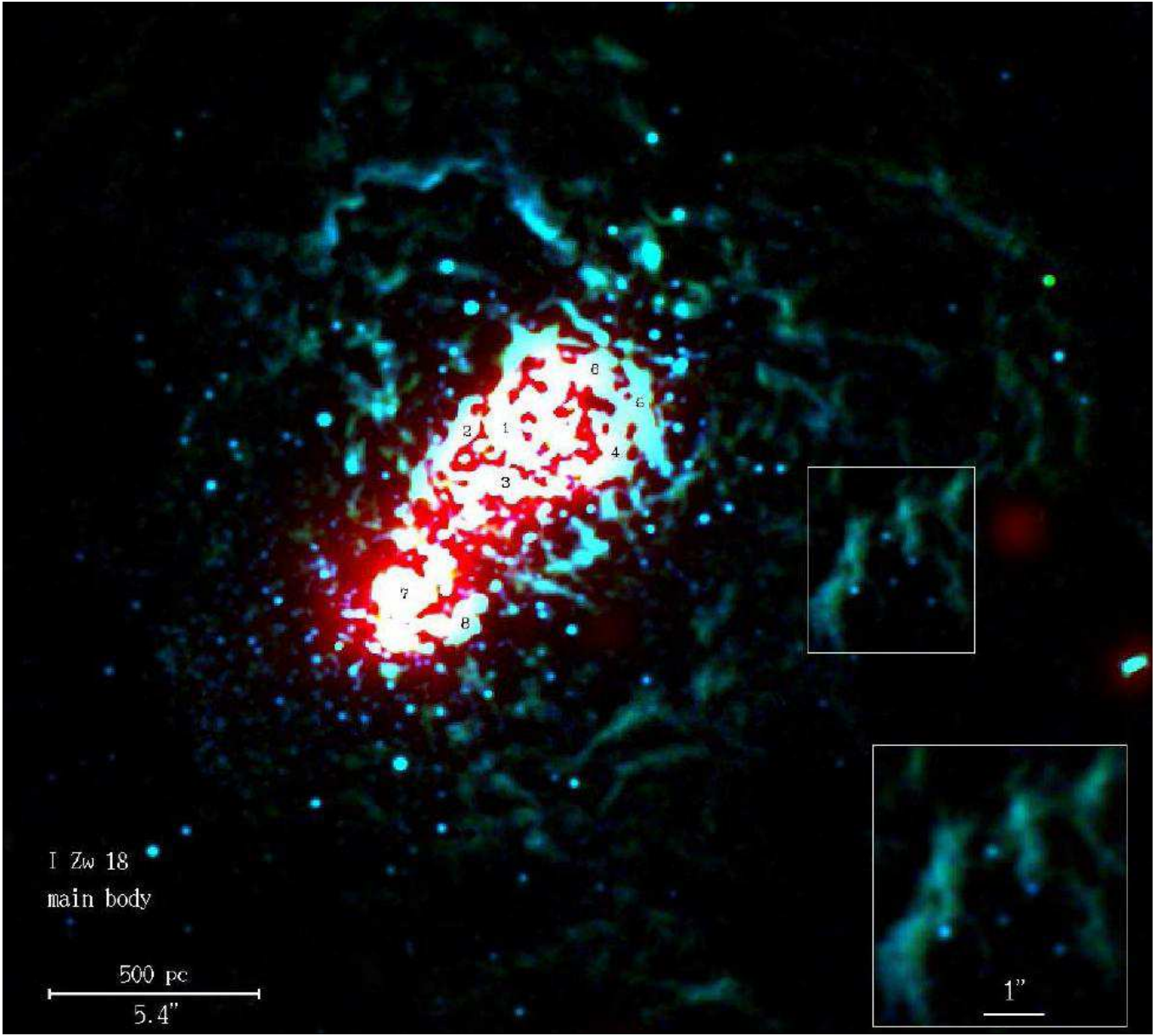,width=18.0cm,angle=0,clip=}}}
\end{picture}
\caption[]{
Three-color image of \iz18\ displaying the unresolved stellar
  emission ($I_{\rm d}$) in the red channel and the $R$ and $V$ emission of 
compact \ige-dominated regions in the green and blue channel.
White areas in the central (red) part of the galaxy depict regions whose 
colors are strongly affected by nebular emission, as apparent from their extremely 
blue $V$--$I$ and $R$--$I$ ($<$ --0.4 \dots --0.8) colors.
The inset, showing a magnified portion of the western super-shell, is meant to
illustrate a subtle spatial shift between the $V_{\rm c}$ and $R_{\rm c}$
emission. This can be attributed to a spatial displacement 
by 30 -- 100 pc between the intensity maxima of the 
[O{\sc iii}]$\lambda\lambda$4959,5007 and \ha\ emission lines.
}
\label{IZw18-ige}
\end{figure*}
\subsubsection{The impact of nebular emission on the colors of the stellar component
of I\ Zw\ 18 \label{kentro}}
We discuss next in more detail the effect of \ige\ contamination in the 
HSB core, i.e. the region subtended by the 25 $R$\arcmin\ \sbb\ isophote
in Fig. \ref{IZw18HaImage}. Our color maps reveal here a substantial substructure 
that was not sufficiently recovered on previous data.
A previously known feature is the \ige\ rim 
\citep[EW(\ha)=1500--2000 $\AA$;][]{Ostlin96,VilchezIglesiasParamo98-IZw18,Papaderos01-IZw18,Izotov01-IZw18} that
encompasses the SF region NW. Its $V$--$I$ and $R$--$I$ colors are much bluer
(--0.8 mag and --0.6 mag, respectively) than those of the centrally located
young SCs (--0.15 mag and --0.23 mag, measured within a 2\farcs8$\times$2\farcs8
aperture) where the EW(\ha) is much lower \citep[$\sim$200 $\AA$;][]{Izotov01-IZw18}.

This striking spatial anti-correlation between stellar surface density and
emission-line EWs, with the $V$--$I$ color steeply decreasing with the increasing 
EW in the periphery of ionizing SCs is essentially identical to that described in the
XBCD \object{SBS 0335-052E} by P98: the bluest $V$--$I$ colors (--0.8 mag) in that
galaxy were not observed at the position of its young SCs but in their periphery,
over an extended horseshoe-shaped gaseous rim some 500 pc offset from the
former. Several similar examples are documented among XBCDs and BCDs
\citep[e.g.][]{Papaderos99-Tol65,Guseva01,Fricke01-Tol1214,Ostlin03-ESO338,Guseva04-Pox186,Papaderos08}
both on small and large spatial scales.

Of considerable interest is also the small-scale contamination of colors
in the HSB core of \iz18\ due to \ige. To better illustrate its impact,
we computed a pseudo three-color composite (Fig. \ref{IZw18-ige}) using a 
combination of unsharp masked images (see discussion in Sect. \ref{phot:IZw18C}).
In order to partly suppress the stellar contribution in
$V_{\rm c}$ and $R_{\rm c}$ images (blue and green channel, respectively), 
we subtracted from the latter a pseudo stellar continuum
using a scaled version of the $I_{\rm c}$ image.
This procedure allowed to better isolate and visualize regions where \ige\ 
contamination is strongest.
The red channel of Fig. \ref{IZw18-ige} holds the unresolved emission 
in the $I$ band ($I_{\rm d}$), it thus primarily reflects the stellar
host galaxy.
In the resulting three-color overlay, white patches on top the reddish 
background of the HSB core depict regions where \ige\ dictates colors.
Just like in radial profiles and color maps
(Figs. \ref{IZw18SBP}, \ref{IZw18-VRImaps} and \ref{IZw18-VImap}), 
the footprint of \ige\ is a combination of red $V$--$R$ with extremely blue $V$--$I$ and $R$--$I$ colors.
Some examples include the regions labeled 1 through 8, whose respective 
colors are (0.1, --0.35, --0.45), (0.17, --0.58, --0.76), (0.05, --0.41, --0.45),
(0.23, --0.43, --0.67), (0.24, --0.45, --0.69), (0.21, --0.53, --0.74),
(0.12, --0.35, --0.47) and (0.23, --0.23, --0.46).
The total area of these severely \ige-contaminated regions of $\approx$30 $\sq\arcsec$
is equivalent to the isophotal size of \iz18\ at 22 \sbb\ (32 $\sq\arcsec$) 
and to 20\% of that of its \emph{stellar} component (the region subtended 
by the 25 $R$\arcmin\ \sbb\ isophote in Fig. \ref{IZw18HaImage}).

\begin{figure}
\begin{picture}(8.6,6.9)
\put(1.3,0){{\psfig{figure=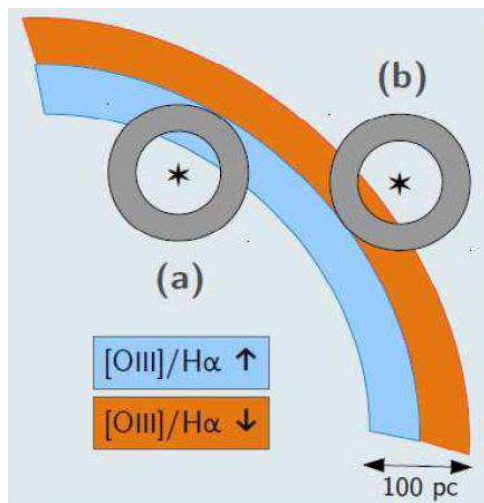,width=6.4cm,angle=0}}}
\end{picture}
\caption[]{Schematic illustration of the effect that the slight spatial
displacement between the [O{\sc iii}]$\lambda\lambda$4959,5007 and 
the \ha\ emission lines across an expanding shell of ionized gas might have
on $V$ and $R$ photometry of stars in its close vicinity. 
The shell, of thickness 100 pc ($\approx$1\arcsec at the distance to \iz18) 
comprises an inner and outer zone with, respectively, a higher and lower 
[O{\sc iii}]/\ha\ ratio (cf the discussion related to
Figs. \ref{IZw18-VRImaps},left and \ref{IZw18-ige}). 
Two cases may be distinguished: in the case of a star located close to the
inner boundary of the shell (a), the local $V$ background within a 
circular annulus will be overestimated due to the enhanced contribution 
of the [O{\sc iii}] lines in the inner shell interface. 
This would result in an artificially red $V$--$R$ color. 
The opposite would be the case for (b) where the local $R$ background 
would be over-subtracted due to the lower [O{\sc iii}]/\ha\ ratio in 
the outer layer of the shell, leading to an artificially blue 
$V$--$R$ color. The principal effect would therefore be that the 
magnitudes of stars in the vicinity of shells would be underestimated by a 
different amount in different bands, with the $V$--$R$ colors of stars 
in the interior of the shell being redder than in its outer part. 
}
\label{CMD_effect}
\end{figure}
It is noteworthy that \ige\ shows significant substructure on spatial scales of 
a few pixels, with little spatial correlation to the local stellar
background. Therefore, its treatment and subtraction as uniform 
foreground emitting layer in CMD studies could result into systematic 
uncertainties that might not be fully accounted for by the standard error budget.
Subtle spatial displacements between the intensity maxima of 
strong nebular emission lines may be of some importance in this regard.  
An example is given in the inset of Fig. \ref{IZw18-ige} where we show magnified 
a portion of the western super-shell. It can be seen that the inner layer 
along the shell appears blueish whereas the outer one greenish. 
This can be plausibly attributed to a displacement by 0\farcs1 -- 0\farcs3
between the [O{\sc iii}]$\lambda\lambda$4959,5007 and \ha\ emission 
lines which are registered in $V$ (blue channel) and $R$ (green channel), respectively.
Another illustrative example is given for a shell $\sim$5\arcsec\ northwards of region NW 
(Fig. \ref{IZw18-VRImaps}) whose inner (southeastern) and outer (northwestern) layer differ by
$\approx$0.3 mag in their $V$--$R$ color, pointing to a variation by
$\ga$30\% in the [O{\sc iii}]/\ha\ ratio across the shell.
Note that systematic variations of the [O\,{\sc iii}]$\lambda$5007/\hb\ ratio 
by a factor $\sim$3 on spatial scales of a few ten pc have been documented by
integral field unit spectroscopy in a number of \h2\ regions and BCDs
\citep[see e.g.][]{Westmoquette07-NGC1569,Cairos09-Mrk409,Relano2010-NGC595,Monreal2010-NGC5253}.

Whereas spatial displacements between nebular lines have certainly 
no measurable effect on CMD studies of galaxies with faint \ige,
they may be of some relevance in the case of \iz18.
This is because the local background level in point source photometry studies
may be overestimated by a different amount in different bands, 
depending on the position of a star relative to a shell and the specifics
of the local background determination.
Figure \ref{CMD_effect} illustrates schematically two special cases:
for a star located near the inner interface of the shell 
(case a) the circular annulus within which the local background is determined
captures the inner [O{\sc iii}]-enhanced portion of the shell
and misses its outer \ha-enhanced layer. 
This could lead to an over-subtraction of the local $V$ background, hence, 
to a reddening of the $V$--$R$ color.
The opposite might be the case for a star close to the outer boundary 
of the shell (case b), whose color would appear bluer due to the over-estimation 
of the $R$ (\ha\ enhanced) background.
In the simple geometry of Fig. \ref{CMD_effect}, the principal effect 
would then be a redder color for stars in the shell interior 
and \emph{vice versa}, in addition to underestimated stellar magnitudes
because of over-subtraction of the local background.
An examination of the cumulative effect that multiple overlapping \ige\ shells
could have on CMD studies of \iz18\ might therefore be of some interest.

Of special relevance to the study of the evolutionary status of \iz18\
(Sect. \ref{age-IZw18}) are the colors of region $\omega$ (Fig. \ref{IZw18-Fig1}).
This region, roughly delimited by the rectangular area at 
(RA,DEC)=(5\farcs8 \dots 4\farcs6,--6\farcs6 \dots --4\farcs4) 
relative to NW (see Figs. \ref{IZw18-VRImaps} and
\ref{IZw18-VImap}) is relatively free of \ige\ contamination
\citep[cf the EW(\ha) map in Fig. 7 of][]{Izotov01-IZw18}, 
its colors can therefore be used to place constraints on the age of \iz18.
From the present data we infer a mean 
$V$--$R$ color of --0.06 mag and a $R$-$I$ and $V$--$I$ color of $\approx$0.2 mag
with values of $V$--$R$ $\simeq$ $R$-$I \approx 0.15$ mag and $V$--$I$=0.3 mag 
within its reddest quarter. The $V$--$I$ color range inferred for region $\omega$
from \hst\ ACS data is in good agreement with the values 0.17 -- 0.32 mag 
previously determined by P02.

\section{Discussion \label{discussion} }

\subsection{Constraints on the evolutionary status of \cc\ 
\label{IZw18C-evol}}
\begin{figure}[h]
\begin{picture}(8.6,8.6)
\put(0.04,4.6){{\psfig{figure=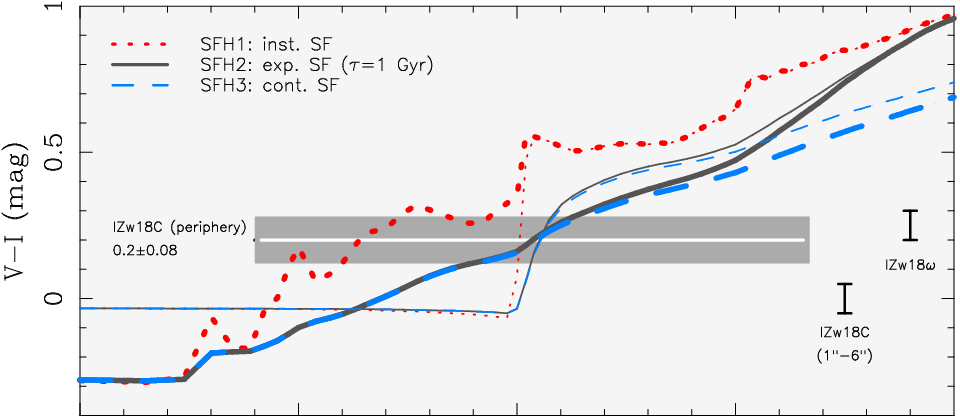,width=8.8cm,angle=-90}}}
\put(0.06,0){{\psfig{figure=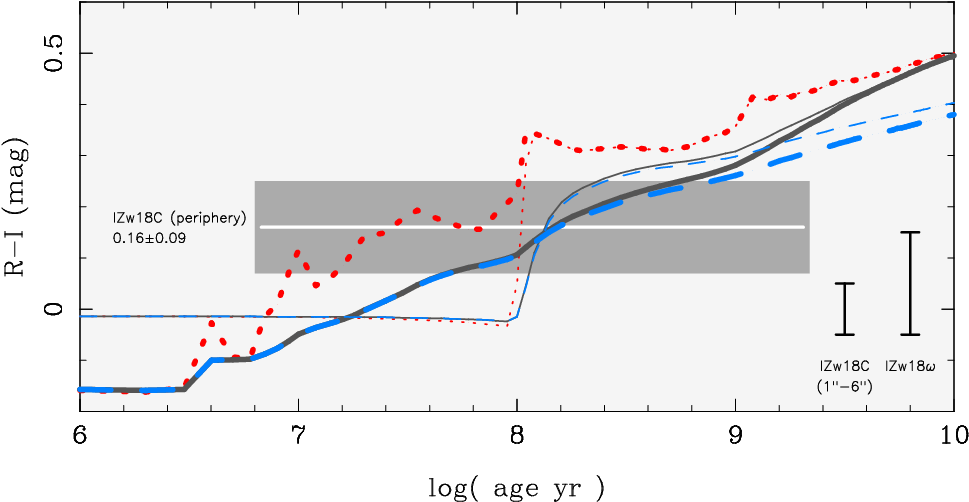,width=8.88cm,angle=-90}}}
\end{picture}
\caption[]{
Comparison of the $V$--$I$ and $R$--$I$ colors 
of \cc\ and of region $\omega$ in \iz18\ with model predictions 
for a stellar population forming instantaneously (\sfha: thick dotted curve) 
or continuously, with an exponentially decreasing or constant 
star formation rate (\sfhb\ and \sfhc: thick solid and dashed line, respectively). 
The models have been computed with Pegase~2.0 \citep{FR97} for
a constant metallicity ($Z$=0.001) and {\bf a}
Salpeter initial mass function between 0.1 and 100 \msun, and 
do not include nebular emission.
Thin curves correspond to the same star formation histories but
assume an IMF truncated above 5 \msun.
The vertical bars labeled I\,Zw\,18$\omega$ depict the range between the
mean and reddest color in region
$\omega$ of \iz18\ ($V$--$I$=0.2 \dots 0.3 and $R$--$I$=--0.06 \dots 0.15).
Bars labeled \cc\ (1\arcsec--6\arcsec) indicate
the mean color of \cc\ in the radius range 1\arcsec$\leq$\rr$\leq$6\arcsec\ 
($\approx$0$\pm$0.05 mag). 
The mean colors and their 1$\sigma$ uncertainties in 
the extreme periphery (\rr$\geq$6\arcsec) of \cc\ at $\mu\ga27.6$ \sbb\ 
are depicted by the rectangular areas.
}
\label{IZw18-pegase}
\end{figure}

In Fig. \ref{IZw18-pegase} we compare the observed colors of \cc\ 
with the predicted color evolution for a stellar population forming
instantaneously (\sfha, dotted line), with an exponentially decreasing 
star formation rate (SFR) and an e-folding time $\tau=1$ Gyr (\sfhb, solid
line) and continuously with a constant SFR (\sfhc, dashed line). 
The theoretical curves were computed with the evolutionary synthesis code Pegase~2.0
\citep{FR97} for constant stellar metallicity of $Z$=0.001 and Salpeter 
initial mass function (IMF) between 0.1 and 100 \msun, and do not include nebular emission.

As already pointed out in Sect. \ref{phot:IZw18C}, the blue and 
nearly constant (0$\pm$0.05 mag) $V$--$I$ and $R$--$I$ colors of 
\cc\ down to 27.6 \sbb\ imply that the photometrically dominant stellar 
population in this system is almost uniformly young.
Formal upper age estimates within 1$\sigma$ uncertainties 
range from $t_1=15$ Myr for \sfha\ to $t_2\sim$35 Myr for \sfhe.
The latter SF scenarios are, however, incompatible with the data
since star formation starting at $t_2$ and continuing to the present 
would have given rise to amble \ige\ with an EW(\ha)$\ga$800 $\AA$.
The uniformly blue colors of \cc, in connection with the absence of 
\ige\ everywhere but in its central region indicate that SF activities 
must have ceased very recently, some $\sim$20 Myr ($= t_{\rm c}$) ago. 
If so, the estimated SF duration ($t_2 - t_{\rm c}\simeq t_1$) would imply
an almost instantaneous SF process, rapidly synchronized over the projected area 
of the galaxy ($\sim$1 kpc$^2$) at probably supersonic speeds.

A conceivable scenario might invoke sequential star formation from the
redder northwestern tip towards the bluer southeastern tip of 
\cc\ \citep[see also][]{Aloisi99-IZw18,Izotov01-IZw18}. 
The respective colors of those regions translate by \sfha\ to an 
age difference of $\tau_{\rm SF} = 40 - 80$ Myr. 
Taking $\tau_{\rm SF} \sim 60$ Myr as an indicative time scale for the spatial
progression of SF activities along the projected major axis of \cc\ (1 kpc), one can estimate
an average SF propagation velocity of $u_{\rm SF} \sim 20$ \kmsec.
The latter is comparable to the $u_{\rm SF}$ inferred for the 
XBCD \object{SBS 0335-052E} 
\citep[$\sim$ 20 -- 35 \kmsec, P98,][]{Reines08-SBS0335},
of the order of the sound speed in the warm ISM.
Induced gas collapse along the collisional interface of a super-shell expanding 
from the northwestern tip of \cc\ between 80 Myr ($\tau_{\rm SF}$+$t_c$) and $t_c$ ago
might offer a tenable, though not necessarily unique explanation 
for propagating star formation: following \cite{McCrayKafatos87}, 
the radius $R_{\rm sh}$ (pc) of a SF-driven super-shell can be approximated as
\begin{equation}
R_{\rm sh} = 269\,\left( \frac{L_{\rm m,38}}{n_0} \right)^{1/5}\, t_7^{3/5}
\label{eq:MC}
\end{equation}
where $L_{\rm m,38}$ is the mechanical luminosity injected into the ISM
by stellar winds and SNe in $10^{38}$ erg\,s$^{-1}$, 
$n_0$ the ambient gas density in cm$^{-3}$ and 
$t_7$ the dynamical expansion time in $10^7$ yr. 
A rough estimate on the mean mechanical luminosity 
can be inferred from the absolute $V$ magnitude of \cc\ (--12.2 mag)
which translates for continuous star formation over $\tau_{\rm SF}$ and an
ensuing quiescent phase over the past $t_c$ to a stellar mass of 
$\approx 2.1\times 10^5$ \msun\ and a mean SFR of $3.5\times 10^{-3}$ \msun\ yr$^{-1}$.
The mean $L_{\rm m,38}$ over $\tau_{\rm SF}$ may be estimated from 
Starburst99 \citep{Starburst99} to be 14.6.
Equation \ref{eq:MC} then yields for $R_{\rm sh}=1$ kpc and 
$t_7=6$ ($\equiv\tau_{\rm SF}$), an ambient gas density $n_0 \approx 4$ cm$^{-3}$.
This value does not seem to be unrealistic, given that the VLA HI map by
\cite{vanZee98-IZw18} indicates for \cc\ a H{\sc i} surface density of 
$\ga 4\times 10^{20}$ cm$^{-2}$, which after inclusion of the helium contribution
translates for a disk thickness of $\sim$50 pc to $n_0 \sim 3$ cm$^{-3}$.
As the passage of the SF front along \cc\ has likely been accompanied by the 
depletion of its molecular content and the dispersal of its ISM, the present 
gas surface density should rather represent a lower estimate on the average
$n_0$ over  $\tau_{\rm SF}$.
Clearly, the considerations above rely on strongly simplifying assumptions 
(constant $L_{\rm m,38}$ and $n_0$, no radiative losses in the shell, 
coplanar face-on geometry) and are meant to be indicative only.
They merely touch upon one among several possible scenarios 
behind propagating star formation in \cc\ (e.g., gravitational
interaction with the main body or gas infall from the H{\sc i} halo) 
and are invoked to demonstrate that this process can account for some
important characteristics of the galaxy.
These include its uniformly blue colors with merely a weak color 
gradient along its major axis, the virtual absence of \ige\ 
and the higher surface density of the unresolved stellar background in its 
northwestern half as possible signature of the disintegration of its oldest SCs.
However, propagating star formation between 
$\tau_{\rm SF}$+$t_c$ and $t_c$ can alone not explain the slightly 
redder color of stellar populations in the outskirts of \cc.

\subsubsection{Stellar diffusion and its effect on age determinations
in the faint periphery of \cc\ \label{mass-filtering}}
In the following we further explore the evolutionary history of \cc\ 
by considering the properties of its faint (27.6 -- 29 \sbb) stellar periphery.
The ($V$--$I$)$_{\rm d}$ (0.2$\pm$0.08) and $R$--$I$ (0.16$\pm$0.09) color of
the latter translate by \sfhb\ to an age of $\sim$130 Myr 
with an upper bound of $\sim$500 Myr at the 1$\sigma$ 
level, if purely stellar emission is assumed.
For models including \ige, age estimates would rise to 270--900 Myr, 
depending on the color considered. 
However, even for those high ages, this SFH model (and \sfhc\ alike) 
predicts an EW(\ha) between 150 $\AA$ and $\ga$300 $\AA$ 
in clear conflict with the absence of \ige\ in the outskirts of \cc.
Models invoking continuous star formation to the present are thus 
fundamentally incompatible to the data.
Cessation of SF activities over the past $t_c$ would alleviate the contradiction, 
it would, however, at the same time imply a steeper color evolution, thus a 
younger age, in better agreement with the 1$\sigma$ upper estimate 
$t_{\rm SFH1} \sim 100$ Myr read off Fig. \ref{IZw18-pegase} for the 
instantaneous star formation model.

As we shall argue next, stellar diffusion and the resulting radial 
\emph{stellar mass filtering effect} described by P02 could have a 
significant effect on the color distribution of young galaxy candidates and 
add an important element towards a consistent evolutionary picture for \cc.
Already at a mean radial velocity of $u_{\rm r} \la 4$ \kmsec, 
less than the velocity dispersion of the neutral gas 
($\sigma_{\rm HI}\approx$6--8 \kmsec), a star born in the central part of \cc\ could migrate within 
$\tau_{\rm diff}=\tau_{\rm SF}+t_c$ (80 Myr) out to $r_{\rm diff} \approx 300$ pc from its initial locus.
This galactocentric distance corresponds to the semi-minor axis of \cc\ at 29 \sbb,
i.e. it is topologically identifiable with the redder, outer exponential SBP feature 
in Fig. \ref{IZw18C_sbp}.
A consequence of stellar diffusion with the above radial velocity
is that any stellar generation reaching out to $r_{\rm diff}$ can not be bluer than 
0.3 mag in $V$--$I$ (approximately the 1$\sigma$ color bound, depicted by
the rectangular area in the upper panel of Fig. \ref{IZw18-pegase}), 
since on its way to that radius it will have been de-populated from stars 
with lifetimes $\leq \tau_{\rm diff}$
(or, correspondingly, from stars more massive than $M_{\rm diff}$).  
The commonly adopted \sfhe\ models would then inevitably result in an overestimation of 
the stellar age in the LSB periphery (\rr$\simeq$$r_{\rm diff}$) 
of young galaxy candidates, such as \cc.
This is because such SFH parametrizations are throughout applied on the implicit
assumption that stellar populations form in the radial zone where they are
currently observed, hence they invariably contain a certain 
fraction of massive blue stars younger than $\tau_{\rm diff}$. 
Obviously, a match between predicted and observed colors is then only 
possible for an age $t_{\rm SFH2,3}$ exceeding the true stellar 
age $t_{\star}$ at the radius \rr, i.e. the condition
\begin{equation}
 \tau_{\rm diff} \leq t_{\rm SFH1} \leq t_{\star} < t_{\rm SFH2,3}
\end{equation}
applies throughout, in particular when models including \ige\ are adopted.

That, in the presence of stellar diffusion, the age of the LSB host of 
young BCD candidates may be overestimated when continuous star
formation models are used is especially apparent 
when projection effects are additionally taken into account: 
in a spherical-symmetric geometry, the colors registered at  
\rr$\equiv r_{\rm diff}$ reflect the luminosity-weighted average 
of stars at radii $\geq r_{\rm diff}$, since the observer's 
line of sight crosses more remote foreground and background 
galaxy zones of longer $\tau_{\rm diff}$. 
Therefore, even though \sfhe\ may provide a reasonable 
approximation to the \emph{integral} SFH of some star-forming
dwarf galaxies (SFDGs), it should not be taken for granted that they
are universally applicable to \emph{spatially resolved} age-dating studies.
This is already suggested by the failure of e.g. \sfhc\ to reproduce 
the generally red colors ($V$--$I$=0.9, $B$--$R$=1.1) of the host galaxy of
typical BCDs (see e.g. P02), even for ages exceeding the Hubble time.

The effect of stellar diffusion to the radius $r_{\rm diff}$ 
is equivalent to in situ star formation at $r_{\rm diff}$ with an 
IMF truncated at masses above a radially decreasing limit  
$M_{\rm diff}$($\tau_{\rm diff}$) (hereafter tIMF).
To illustrate this point, and as a minimum consistency check, one may consider
continuous star formation according to \sfhe\ with a tIMF limited 
to stars with a lifetime $\ga \tau_{\rm diff}$. 
This timescale roughly corresponds to the main sequence lifetime of 
a B4 star (94 Myr), placing an upper cutoff of 5 \msun\ to the tIMF.
The color evolution for \sfhe\ with that tIMF is depicted in
Fig.~\ref{IZw18-pegase} with thin lines.
It can be seen that \sfhe+tIMF models reproduce the observed colors 
at an age of $\sim$130 Myr with an 1$\sigma$ upper bound of
150--250 Myr and consistently account for the absence of nebular emission.

A conceivable alternative to stellar diffusion is in situ star formation 
in the LSB periphery of \cc\ in conjunction with stochastic effects 
on the IMF \citep[see][and references therein]{CervinoMasHesse94}.
The latter are to be expected when low masses are involved in the SF process,
as is typically the case for SFDGs, and to affect the sampling of the IMF in
its intermediate-to-high mass range.
Cervi\~no \& Mas-Hesse (see their Fig.~1) argue, however, that stochastic effects 
do not result in a truncated IMF above a certain stellar mass, but
some massive stars always form.
On the statistical average, one would therefore expect nebular emission 
to be present in the outskirts of \cc, in disagreement with the observations.
It should be noted, though, that this conclusion is likely dependent on the detailed physics 
of star formation, from which the high-mass end of the IMF is populated 
and the time scales for low and high mass star formation.
With certain assumptions \citep[see e.g.][]{WeidnerKroupa2005,Pflamm-Altenburg2009} 
a truncated IMF can be realized in peripheral galaxy zones, lifting the discrepancy described above.
There might therefore be variants of the traditional {\sl static} scenarios, i.e. scenarios 
implicitly assuming that stars are born in the radial zone where they are observed,
that can quantitatively reproduce the radial color distribution of \cc\ and SFDGs in general.
It is not clear, however, whether such static scenarios, and their assumptions on the radial 
dependence of the IMF, can consistently account for the specific structural
characteristics of SFDGs, and in particular BCDs, such as e.g. the exponentiality of the stellar LSB host and 
the confinement of star-forming activities to within \rr$\approx 2\alpha$ (see e.g. P96b).

Stellar diffusion and its role on the buildup of SFDGs constitutes an unexplored territory in dwarf galaxy research. 
A caveat of the above discussion is that it assumes young stars to
initially drift apart freely at a roughly constant velocity 
$u_r$ ($\la\sigma_{\rm HI}$) over $\sim 10^8$ yr. 
Clearly, the validity of this simplifying assumption needs to be investigated,
and the initial kinematics of the newly formed stars as well as the form of
the gravitational potential be included in the analysis.
High-velocity resolution ($\la 10$ \kmsec) integral field
spectroscopy might provide key insights into the velocity 
patterns and the possible outwards migration of bound and unbound
stellar clusters in young starbursts. 
A closer investigation of the effects of diffusion would be of 
considerable interest, among others because it offers a mechanism that naturally 
produces radial color gradients in galaxies. 
As such, it may be of relevance for a range of topics, such as e.g.
the \emph{Red Halo} phenomenon \citep{BergvallOstlin02,Z06,BZC10}.

In summary, our results suggest that the overall 
photometric properties of \cc\ can be consistently accounted for by 
a single episode involving solely sequential star formation 
($u_{\rm SF} \approx 20$ \kmsec) and stellar diffusion 
($u_{\rm r} \approx 4$ \kmsec) over the past $\sim$100--200 Myr.
Whereas an ultra-faint substrate of ancient stars can not \emph{per se}
be ruled out, the lines of evidence presented here indicate
that \cc\ is a cosmologically young object.
This conclusion concurs with that by \cite{Izotov01-IZw18} who inferred an 
upper age of $\sim$100 Myr for \cc\ from spectral synthesis models.
\cite{Jamet10-IZw18} reached a similar conclusion, that \cc\ is younger than
$\approx$125 Myr, from a probabilistic analysis of \hst\ ACS CMDs.
 
\subsection{Constraints on the evolutionary status of \iz18\ \label{age-IZw18}}
We turn next to the evolutionary status of \iz18's main body.  
The $V$--$I$ and $R$--$I$ color range in region $\omega$
(Fig. \ref{IZw18-pegase}), translate by \sfhe\ to an age between
$\sim$100 and $\la$250 Myr. 
The possibility of \iz18\ and \cc\ forming a co-evolving pair of dwarf galaxies
that underwent a roughly synchronous strong recent evolution can not be ruled out.

By considering the reddest quartile of region $\omega$, one obtains from \sfhe\ 
an upper age of $\sim$500 Myr. The age span inferred from the present study
is consistent with that obtained from the mean $B$\arcmin--$V$\arcmin\ (0.09$\pm$0.04 mag) 
and $V$\arcmin-$R$\arcmin\ (0.12$\pm$0.04 mag) for the stellar host
galaxy of \iz18\ after two-dimensional subtraction of strong nebular lines
from broad band images (P02).
Note, however, that the latter emission-line free colors are likely slightly
overestimated, as they do not include corrections for the red \citep[$B$--$V$=0.34,
$V$--$R$=0.64, see e.g.][]{Krueger95} nebular \emph{continuum}.
Notwithstanding this fact, even when taken within their +1$\sigma$ bounds 
at face value, they imply by \sfhb\ an age of $\sim$0.8 Gyr, translating 
into a mass fraction of $\ga$50\% for stars younger than 0.5 Gyr.
Evidently, this conclusion \citep[see also][]{Hunt03-IZw18} is not in conflict
with the presence of a small number ($\la$20) of stars with ages between 0.5
and $\sim$1 Gyr in CMDs \citep{Aloisi99-IZw18,Ostlin00,OstlinMouhcine05}. 
On the other hand, it is important to bear in mind that estimates on the mass 
fraction of young stars depend critically on the adopted SFH. 

\subsubsection{An extended underlying stellar disc in I Zw 18?\label{izw18-disk}}
We next revisit the hypothesis envisaged by \cite{Legrand00-IZw18}, 
according to which the formation of \iz18\ is occurring throughout its H{\sc i}
halo (60\arcsec$\times$45\arcsec) at an extremely low SFR over the past $\sim$14 Gyr.
This model can reconcile the low gas-phase metallicity of \iz18\ with 
a cosmological age, and predicts a high degree of chemical homogeneity in its
warm ISM. 
From the photometric point of view, the main prediction from the
Legrand model is an extended ultra-LSB underlying stellar disk 
($\overline{\mu}\simeq$28 $V$ \sbb).
In a subsequent study, \cite{Legrand01-IZw18} have generalized the scenario of
'slowly cooking' dwarfs to BCDs, arguing that their main 
metallicity and stellar mass contributor is continuous star
formation, rather than coeval starbursts.

The hypothesis of a stellar disk beneath the nebular envelope of \iz18\ was
later on investigated by P02 on the basis of emission-line free \hst\ WFPC2 
$B$\arcmin, $V$\arcmin\ and $R$\arcmin\ images.
By consideration of photometric uncertainties, P02 argued that the predicted
disk would evade detection on their SBPs if its central surface brightness 
$\mu_{\rm E,0}$ is fainter than 27.1 \sbb\ and its exponential scale length
$\alpha$ larger than 10\farcs5. These limits translate to an apparent
magnitude of 20 mag and a $\overline{\mu}\simeq$28.6 \sbb, 
slightly fainter, but probably still consistent, within the uncertainties, 
with the $\overline{\mu}$ predicted by the Legrand model.
Thus, previous surface photometry could neither strictly rule out nor establish the 
presence of the putative ultra-LSB disk in \iz18.
Since deep \hst\ ACS narrow band imagery is unavailable for \iz18, we can
not improve on the nebular line subtraction carried out by P02 and push previous 
$V$\arcmin and $R$\arcmin\ surface photometry to fainter levels.

However, CMD studies of \iz18\ argue against the disk hypothesis, as
none among them has revealed a uniform and extended population of RGB 
stars being cospatial with the nebular envelope, 
at sharp contrast to any evolved BCD studied as yet. 
Admittedly, as none among the published CMD studies for \iz18\ 
fully appreciates the importance of \ige\ contamination (except 
for that by IT04) and attempts a proper assessment of the systematic 
errors this can introduce, the robustness of the non-detection 
of RGB candidates in the LSB envelope can not be currently evaluated, 
neither is it clear whether CMD studies can ever place firm constraints in this respect.
The currently most convincing evidence for the absence of a substantial stellar background 
beneath the nebular envelope of \iz18\ is based on the emission-line 
free \hst\ images by P02 (see their discussion).
Note that, even if an ultra-LSB stellar disk complying with the photometric 
limits placed by P02 was present, it would likely not significantly alter the youth
interpretation for \iz18:
The absolute $V$ magnitude of the predicted disk (--11.4 mag), implies for continuous 
SFR over 14 Gyr a stellar mass of $\sim 5.6 \times 10^6$ \msun. 
The $V$\arcmin\ magnitude of the \emph{host} galaxy of \iz18\ (i.e. excluding
regions NW and SE) of --13.8 mag (P02) translates by \sfhb\ for an age between 
0.5 and 1 Gyr to a stellar mass of $\sim 11 \times 10^6$ \msun, i.e. about 
twice that of the hypothetical ultra-LSB disk.
Even in the extreme case of a 14 Gyr old instantaneous burst, the underlying disk would only be 
twice as heavy as the young population which have formed on a much shorter time scale.
Consequently, the conclusion that \iz18\ has formed most of its stellar mass
at a late cosmic epoch still holds.
It would be interesting to check whether the combined chemical output 
from these two stellar populations is consistent with the observed
gas-phase metallicity in \iz18.

With regard to the proposed generalization of the 'slowly cooking' scenario, 
it should be noted that the Legrand model faces difficulties in reproducing basic
photometric properties of BCDs. 
Specifically, the host galaxy of typical BCDs is incompatible with an
ultra-LSB disk and, to the contrary, its central stellar density 
($\geq$1 \msun\ pc$^{-3}$) exceeds by an order 
of magnitude that of dwarf irregulars or dwarf spheroidals (P96b).
Similarly, the mean surface brightness of the BCD host is 
$\leq$24 $B$ \sbb\ \citep[e.g. P96b,][]{Amorin09-BCDs}, roughly 4 mag brighter 
than the model prediction. 
Additionally, as pointed out in Sect. \ref{mass-filtering}, continuous star
formation models are inconsistent with the red host galaxy colors of most nearby BCDs.
In summary, whereas appealing from the chemical point of view, the 'slowly cooking'
scenario does not seem to be reconcilable with the structural properties of
BCDs. This scenario might, however, provide a good approximation to
quiescent late-type dwarfs, such as e.g.
{\sl blue low surface-brightness galaxies} \citep{RB94}.

\subsection{I Zw 18 as local morphological template for rapidly assembling 
galaxies at high redshift \label{iz18-z}}
The integral photometric properties of the overwhelming majority of SF
galaxies in the local universe are barely affected by \ige.
Typically, the \ewhn\ ranges between a few ten $\AA$ 
\citep[e.g.][]{MoustakasKennicutt06,KoopmannKenney06} in normal late-type galaxies
to $\la 10^2$ $\AA$ for the majority of local SFDGs
\citep[see e.g.][]{Lee07,SanchezAlmeida08}. 
Even for BCDs, \ige\ has a noticeable photometric impact
\citep[\ewhn$\simeq$ 200 -- 500~$\AA$;][among others]{Terlevich91,Cairos02,BergvallOstlin02,GildePaz03-BCDs,Guseva09-LZ}
in their centrally confined SF component only (i.e. for \rr$\la$\rsf, with
\rsf\ being of the order of $\sim$1 kpc) and is practically negligible for larger 
radii where the evolved stellar LSB host entirely dominates the line-of-sight
intensity \citep[P02,][]{Knollmann04}. 
More generally, notwithstanding the fact that \ige\ in local SFDGs may
protrude beyond \rsf\ 
\citep[][among others]{HunterGallagher1985,Gallagher1989,HunterGallagher1990,
HunterGallagher1992,Meurer92,Marlowe95,Ferguson96-Ha,PapaderosFricke98a,
Martin98,Bomans02,Cairos02}, it has due to its extremely low surface
brightness no impact on surface photometry \citep[P02,][]{Knollmann04}. 
As an example, even for BCDs with strong ongoing SF activity, corrections of 
integral photometric quantities for \ige\ do not exceed $\approx$0.1 mag \citep{Salzer89}.

In the nearby universe, the only cases of SFDGs with extreme \ige\ contamination
are documented in a few XBCDs. 
Intense and almost galaxy-wide SF activity in these rare young galaxy candidates, 
in combination with the low surface density of their underlying stellar host,
boost [O{\sc iii}]$\lambda$5007 and \ha\ emission line EWs to values
of up to $\ga 2\times 10^3$ $\AA$ and $\ga 1.6\times 10^3$ $\AA$, respectively 
\citep[e.g. P98,][among
  others]{Izotov97-SBS0335,Guseva04-Pox186,Izotov06-SBS0335-GIRAFE,Papaderos06-6dF,Papaderos08},
i.e. of the order of the effective width of broad band filters.
Similar, though less extreme, cases are some of the recently discovered 
ultra-compact starbursting dwarfs in galaxy clusters \citep[EW(\ha)$\sim 10^3$ $\AA$,][]{Reverte07}.

\begin{figure}[h]
\begin{picture}(8.2,7.3)
\put(0.3,0.){{\psfig{figure=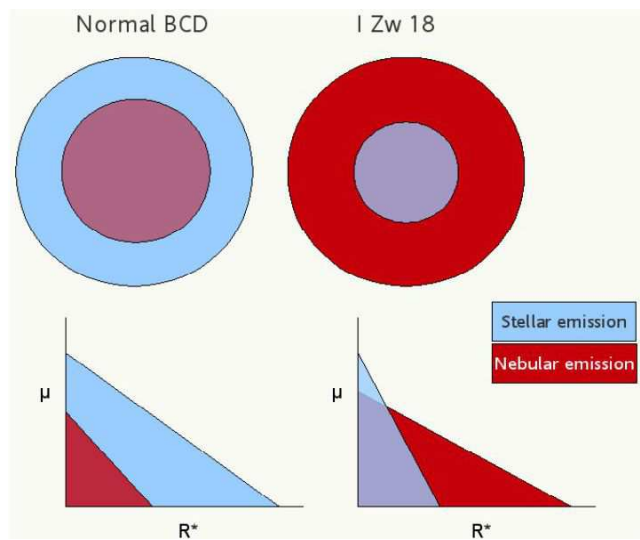,width=8.4cm,angle=0,clip=}}}
\end{picture}
\caption[]{
Schematic comparison of typical Blue Compact Dwarf (BCD) galaxies 
(left) with \iz18\ (right) with respect to the morphology and 
radial intensity distribution (upper and lower figure, respectively)
of their stellar and nebular emission. 
In typical BCDs (and star-forming dwarf galaxies in general) the contribution 
of nebular emission to the line-of-sight intensity is practically negligible, 
especially outside their centrally confined SF component (i.e. for \rr$\geq$\rsf). 
By contrast, in \iz18\ nebular emission dominates already in the inner part of the galaxy
(\rr$\simeq$3\reff, corresponding to a surface brightness level of 23.5 \sbb) 
and produces its \emph{exponential} lower-surface brightness envelope that
reaches as far out as $\sim$16 stellar exponential scale lengths.
}
\label{fig:IZw18_ce}
\end{figure}

As shown in P02, the radial \ha\ intensity profiles of BCDs comprise two
characteristic components: a higher-surface brightness core with 
\rr$\simeq$\rsf\ and an outer, roughly exponential LSB envelope with 
$\alpha_{\rm H\alpha}$ in the range 0.1 -- 1 kpc.
Judging from its $\alpha_{\rm H\alpha}$ (210 -- 270 pc) and radial extent (\rr$\sim$2.6 kpc),
the \ige\ halo of \iz18\ is by no means exceptional among BCDs/SFDGs.
However, \iz18\ strikingly differs from any SFDG studied as yet by the fact 
that the galaxy itself (i.e. its stellar host, cf. Fig. \ref{IZw18HaImage})
is several times more compact than the nebular halo:
\ige\ dominates already for \rr$\simeq$6\arcsec\ ($\equiv$3\,\reff) and reaches as far out as 
$\sim$16 \emph{stellar} exponential scale lengths (Sect. \ref{phot:IZw18}).
This, so far unique, case in the nearby universe is schematically illustrated in
Fig. \ref{fig:IZw18_ce} where the radial distribution of stars and \ige\ in
\iz18\ is compared with that of normal BCDs.

For the forthcoming discussion it is of special importance to recall that the
surface brightness level at which in \iz18\ \ige\ dominates is quite high
($\mu \simeq 23.5$ \sbb; cf Fig. \ref{IZw18SBP}), i.e. comparable to the 
central surface brightness of dwarf irregulars
\citep[e.g.][]{PattersonThuan96,vanZee00-dI} and dwarf ellipticals 
\citep{BinggeliCameron91,BinggeliCameron93}, and by at least one mag brighter 
than that of Local Group dwarf spheroidals \citep[cf.][]{Mateo98-LG}. 
The \ige\ envelope of \iz18\ is therefore not to be confused with the 
extraordinarily diffuse \ige\ in typical SFDGs. 
Of special relevance is as well the fact that the exponential \ige\ envelope 
contributes at least 1/3 of the total $R$ band luminosity of \iz18\ 
(P02 and Sect. \ref{phot:IZw18}).

As we shall argue next, the morphological properties of \iz18, whereas unique
among nearby SFDGs, are likely typical for rapidly assembling galaxies in the distant universe. 
Just like \iz18, these systems are building up their stellar component 
through starbursts or prolonged phases of strongly elevated specific SFR (SSFR),
translating into short (a few 100 Myr) stellar mass doubling times. 
The cumulative output of energy and momentum from stellar winds and SNe during such 
dominant phases of galaxy evolution will inevitably lead to a large-scale 
gas thermalization and acceleration, with super-shells protruding much beyond the galaxy itself.
Extended nebular halos encompassing the still compact stellar component 
of high-SSFR proto-galactic systems may thus be ubiquitous in the early 
universe. The photometric structure of these galaxies could then closely resemble the right
panel of Fig. \ref{fig:IZw18_ce}, comprising a HSB core to within which 
the stellar component is confined and dominates and a much larger, 
nearly exponential, nebular LSB envelope. 
Despite cosmological dimming, the latter should be readily accessible to
observations out to $z \ga 2$, given that deep \hst\ imaging and image
stacking now permit studies of galaxies down to rest-frame surface
brightnesses of $\ga$ 26.5 -- 28.5 \sbb\ at those redshifts
\citep[e.g.][]{Stockton08,vanDokkum2010,Noeske06-UDF}.
Morphological analogs to \iz18\ may as well exist 
among compact high-SSFR galaxies at intermediate redshift (0.1$\la z \la$0.8),
such as e.g. Compact Narrow Emission-Line Galaxies \citep[CNLEGs,][]{Koo94,Guzman98}, 
Luminous Compact Blue Galaxies \citep{Guzman03-LCBG,Puech06-LCBs} and 
Green Pea (GP) galaxies \citep{Cardamone09-GP,Amorin10-GP,Izotov11-GP}. 
Note, that the rest-frame EWs of GPs can in some cases almost compete with those
of nearby XBCDs \citep[see e.g.,][]{Izotov01-SBS335,Papaderos06-6dF}.

\smallskip
Such considerations motivate a closer examination of observational biases 
that the \emph{spatial segregation} between stellar and nebular emission
may introduce into studies of distant, poorly resolved morphological analogs of \iz18. 
By appreciation of the differing spatial distribution of stellar and nebular emission
the discussion here goes beyond the framework of state-of-art modeling studies
\citep[e.g.][among others]{Huchra77,Bergvall85-ESO338-IG04,
Salzer89,Olofsson89,Krueger95,FR97,Izotov97-SBS0335,
Guseva01,Moy01,Anders04-NGC1569,
Zackrisson01,Panuzzo03,Zackrisson08,Kotulla09-GALEV,MM10-POPSTAR,ShaererBarros09,
Finlator11,Ono10} exploring in detail the impact of \ige\ on the
\emph{integral} SED of SF galaxies. 
The predictions from such zero-dimensional models are to be treated with some
caution when compared with \emph{spatially resolved} observables (e.g. radial EW and color
profiles) in order to place constraints on the formation history of galaxies.
The minimum prerequisite for this approach to be valid is that nebular emission is
cospatial with the local ionizing and non-ionizing stellar background, or, 
equivalently, that the ionizing Lyman continuum budget is reprocessed into 
nebular emission \emph{on the spot}. 
That this idealized picture is not invariably valid and, actually, a strong 
spatial anti-correlation between emission line EWs and stellar surface density
may evolve over time both on small and large scales has been shown for several
XBCDs \cite[e.g. P98; P02;][]{Papaderos99-Tol65,Izotov01-IZw18,Guseva01,Fricke01-Tol1214}.
In these cases, subtraction of synthetic \ige\ SEDs from observed spectra, as done in P98, 
offers the only viable approach for isolating and age-dating the underlying stellar SED.

A discussion of aperture effects on the luminosity-weighted SEDs of
distant morphological analogs to \iz18\ is beyond the scope of this study.
Here, we will only focus on potential photometric biases. 
A first one, already described in P02, arises from the fact that the 
nebular envelope may, due to its exponentiality and reddish ($B$--$R$ and
$V$--$R$) rest frame colors, be taken as evidence for an evolved stellar disk. 
Automated surface photometry analyses of large 
extragalactic probes, if not including a robust discriminating criterion
between stellar and nebular emission, could therefore be biased towards an 
enhanced frequency of galactic disks already at place at an early cosmic epoch.
The arguably most convincing argument for this concern derives from
\iz18\ itself: if for one of the best studied galaxies in the local universe, 
it took three decades to realize that its exponential LSB envelope is not due 
to a stellar disk but due to \ige, one may rightfully doubt that this 
would be immediately apparent in studies of its poorly resolved morphological 
analogs at high $z$.

\begin{figure}[h]
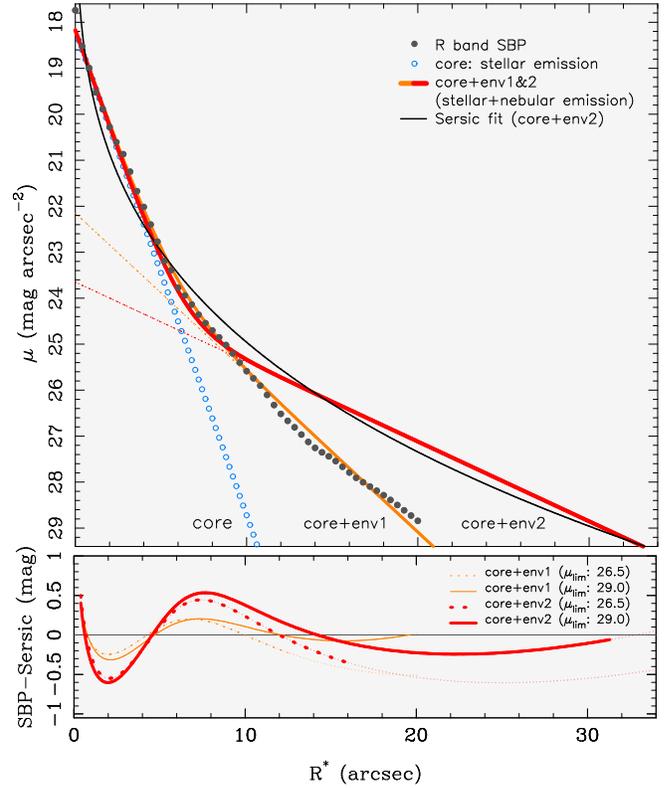

\begin{picture}(8.2,10.6)
\put(0.1,3.1){{\psfig{figure=fig13a.ps,width=8.6cm,angle=-90.0}}}
\put(0.14,0.){{\psfig{figure=fig13b.ps,width=8.54cm,angle=-90.0}}}
\end{picture}
\caption[]{
{\bf upper panel:} Comparison of the $R$ SBP of \iz18\ (filled circles),
with a synthetic SBP (labeled {\nlx core+env1}) that is due to the superposition of
two exponential components of differing central surface brightness
$\mu_0$ and exponential scale length $\alpha$. 
The first one, labeled {\nlx core} (open circles) approximates the stellar component 
that is confined to and dominates within the inner 
(\rr$\leq$6\arcsec) HSB part of the observed SBP. 
The second component ({\nlx env1}; dotted line intersecting 
the abscisca at $\mu \sim 22$ \sbb) is a linear fit to the nebular 
LSB envelope. 
A S\'ersic model to the composite {\nlx core+env1} SBP (orange solid-line curve) 
yields a S\'ersic exponent $\eta \approx $2, close to the best-fitting value of 
$\eta\approx$2.2 for the observed SBP.
The synthetic SBP labeled {\nlx core+env2} 
(red solid-line curve) is computed by superposing on the
{\nlx core} an exponential nebular envelope of equal luminosity but twice 
as large $\alpha$ as the observed envelope {\nlx env1}. 
A S\'ersic fit to {\nlx core+env2} (thin curve) yields an $\eta\approx5$.
{\bf lower panel:} Residuals between the synthetic SBPs {\nlx core+env1} (thin curve)
and {\nlx core+env2} (thick curve) and their S\'ersic fits when the latter are
computed down to a surface brightness level $\mu_{\rm lim}$ of 26.5 and 29
\sbb\ (solid and dotted curve, respectively). 
It can be seen that residuals do not exceed 0.2 mag and 0.5 mag, respectively, 
i.e. they are of the order of or smaller than typical uncertainties in SBPs 
of intermediate-to-high $z$ galaxies at rest frame surface brightness levels
$\mu\la$26.5 \sbb.
}
\label{fig:2exp_SBPs}
\end{figure}

\begin{figure}[h]
\begin{picture}(8.2,5.8)
\put(0.3,0.){{\psfig{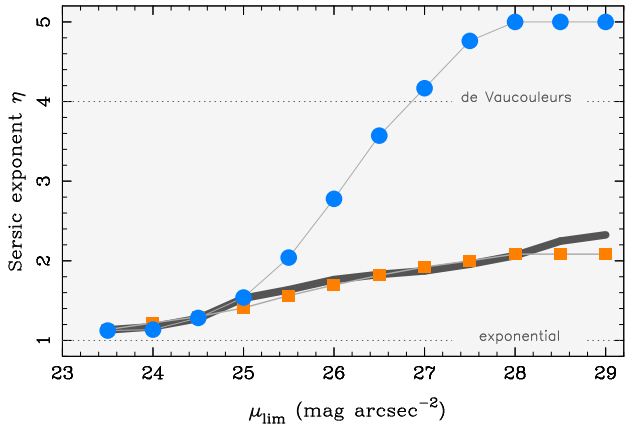}}}
\end{picture}
\caption[]{Best-fitting S\'ersic exponent $\eta$ vs $\mu_{\rm lim}$ for the SBPs
in Fig. \ref{fig:2exp_SBPs}. 
The thick-gray curve and the filled squares show, respectively, the variation of $\eta$ 
for the observed $R$ band SBP of \iz18\ and the two-component approximation to
it, labeled {\nlx core+env1}.
The variation of $\eta$ with $\mu_{\rm eff}$ for the synthetic SBP 
{\nlx core+env2} is shown with filled circles.}
\label{fig:mulim_vs_eta}
\end{figure}

Another potential bias is owing to the fact that the superposition of two exponential
profiles of differing $\mu_0$ and $\alpha$ -- one representing the steeper
star-dominated core and the other the shallower nebular envelope -- can closely
approximate a genuine S\'ersic profile with a high shape parameter
(2$\leq \eta \la 5$), thereby mimicking the SBP of a massive spheroid.
The best-fitting S\'ersic exponent $\eta$ for such a composite SBP depends
both on the properties of its constituent exponential profiles and the limiting 
surface brightness $\mu_{\rm lim}$ down to which S\'ersic models are fitted.
As an example, the best-fitting $\eta$ for \iz18\ increases monotonically 
from 1.1 for $\mu_{\rm lim}=23.5$ \sbb\ (at the surface brightness 
where \ige\ takes over) to 2.2 for $\mu_{\rm lim}=28$ \sbb.
This is illustrated in Fig. \ref{fig:2exp_SBPs}, where the observed SBP of 
\iz18\ (filled circles) is overlaid with the superposition of two
exponential components, approximating the core (open circles; profile labeled
{\llx core}) and the \ige\ envelope (dotted line crossing the abscissa at $\sim$22 \sbb; {\llx env1}). 
Their sum ({\llx core+env1}; orange solid-line curve) is then fitted by S\'ersic
models down to a progressively faint $\mu_{\rm lim}$. 
It can be seen from Fig. \ref{fig:mulim_vs_eta} that the best-fitting
$\eta$ (squares) for the synthetic SBP is doubled when $\mu_{\rm lim}$ 
decreases from 23.5 to $\geq$27.5. 
The same trend is recovered when S\'ersic models are fitted 
directly to the observed SBP (thick curve).

It is worth checking how the S\'ersic $\eta$ vs. $\mu_{\rm eff}$ relation may
change in the case of an equally luminous but shallower nebular envelope.
For this, we superpose to the {\llx core} a component ({\llx env2}) whose $\mu_0$ and
$\alpha$ is by, respectively, 1.5 mag fainter and a factor of 2 larger than that
in the observed component {\llx env1}. 
The S\'ersic fit to the {\llx core+env2} profile (red solid-line curve) is included 
in Fig.~\ref{fig:2exp_SBPs} with the thin--dark curve.
As apparent from Fig. \ref{fig:mulim_vs_eta}, in this case $\eta$ shows a
steeper dependence on $\mu_{\rm lim}$, increasing to $\sim 4$ already for 
$\mu_{\rm lim}\simeq 26.5$ \sbb\ and leveling off to $\sim$5 at fainter levels.  

Consequently, S\'ersic fits can readily lead to the misclassification of a 
compact high-SSFR galaxy as a massive elliptical and, quite counter-intuitively, 
deeper photometry exacerbates the problem.
Note that the residuals between SBP and model (lower panel of Fig. \ref{fig:2exp_SBPs})
are $\la$0.5 mag for {\llx core+env2} and just $\la$0.2 mag for {\llx core+env1}, 
i.e. of the order of 1$\sigma$ uncertainties in SBPs of intermediate-to-high $z$ galaxies \citep[cf
  e.g.][]{Noeske06-UDF}, i.e. small enough to go undetected. 
In practice, even when PSF convolution effects are fully accounted for, 
the pseudo-S\'ersic profile of a compact diskless high-SSFR galaxy is barely
distinguishable from the S\'ersic profile of a massive galaxy spheroid.

Evidently, since extended \ige\ can drastically affect an SBP as a whole, 
it also impacts virtually all secondary photometric parameters that are derivable 
from it (e.g. the effective radius and Gini coefficient, and the various light
concentration indices commonly used in galaxy quantitative morphology studies).

\smallskip
Thirdly, the \ige\ luminosity fraction in \iz18\ ($\geq$1/3), if typical for its 
higher-$z$ analogs, translates into a systematic error of $\geq$0.4 mag 
in galaxy scaling relations involving total magnitudes.
These range from the Tully-Fisher relation to all relations comparing
luminosity with e.g. metallicity, diameter, mean surface brightness and velocity dispersion. 
Moreover, errors in galaxy luminosity propagate, potentially amplified, in 
stellar mass determinations using theoretical mass-to-light ratios or SED fitting.
An investigation of this issue was recently presented by \cite{Izotov11-GP}:
these authors has shown that masses computed from spectral fits to the SED
\emph{continuum} of high-SSFR galaxies at intermediate $z$ can be
overestimated by a factor of up to $\sim$4. 
This is not primarily due to the luminosity contribution but due to the 
red spectral slope of the nebular continuum \citep[see e.g.][]{Krueger95} which 
drives SED fitting towards solutions invoking a much too high
luminosity fraction from old stars.
In galaxy assembly studies covering a wide range in $z$, downsizing
effects may further complicate the aforementioned biases, since \ige\ is expected to 
affect galaxy populations of different mass over different timescales.

\smallskip
We next turn to the color contrast \dc\ between the HSB
\emph{core} and the nebular LSB \emph{envelope}.
This quantity can readily be determined 
from SBPs, or within concentric apertures, and provides a handy proxy 
to radial color gradients in galaxies, thus a first classification guess. 
As we will show below, \dc\ offers for certain $z$ intervals a powerful 
diagnostic for identifying high-SSFR galaxies with morphological properties
analogous to those \iz18. 
Within other $z$ intervals, however, and depending on the colors considered, 
a superficial interpretation of \dc\ can further aggravate the above discussed
galaxy misclassification biases.

In computing \dc\ and its variation with $z$, we approximated the spectrum 
of \iz18\ with synthetic stellar + \ige\ SEDs from Pegase~2, referring to 
\sfhb\ and a metallicity $Z$=0.0004. 
In these models, the properties of the HSB core
(\rr$\leq$6\arcsec) are well reproduced by a synthetic SED for an age $t$=100 Myr. 
This yields colors of $V$--$R$=0.2, 
$V$--$I$\,=\,--0.15, $R$--$I$\,=\,--0.37 mag, and an EW(\ha) of 670 $\AA$, in good 
agreement with the observed values 
\cite[][P02]{Izotov01-IZw18,VilchezIglesiasParamo98-IZw18}.
As for the envelope (6\arcsec$\leq$\rr$\la$20\arcsec), we adopted the SED 
from the same model for $t$=0, i.e. a purely \ige\ spectrum. 
Its colors ($B$--$V$=0.28, $V$--$R$=0.47, $B$--$R$=0.7, $V$--$I$\,=\,--0.65 and 
$R$--$I$\,=\,--1.1 mag) provide as well a good match to the data.

\begin{figure}
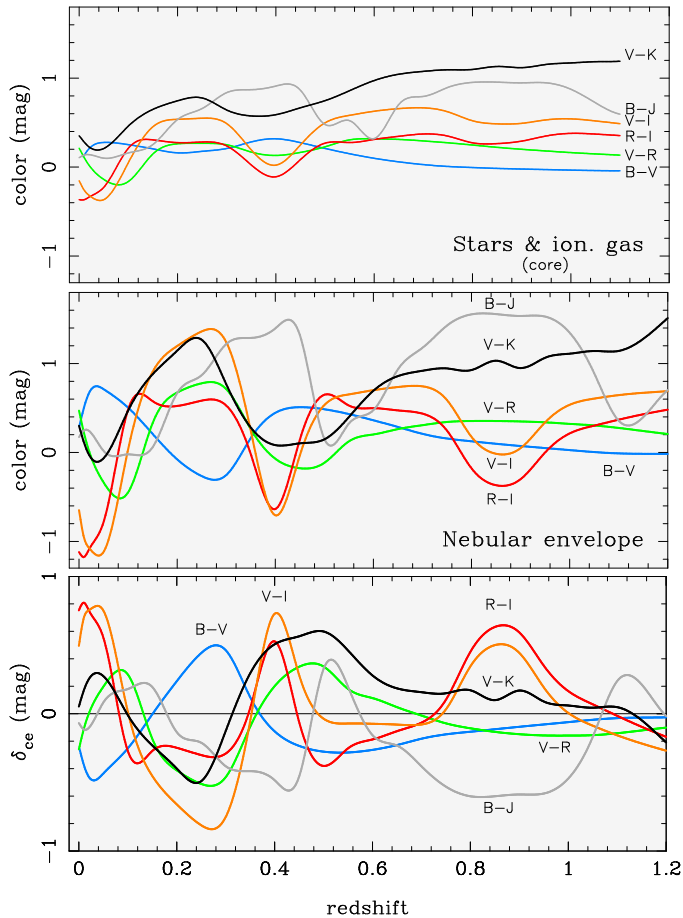

\begin{picture}(8.6,12.36)
\put(0,8.4){{\psfig{figure=fig15a.ps,width=8.8cm,angle=-90.0}}}
\put(0,4.6){{\psfig{figure=fig15b.ps,width=8.8cm,angle=-90.0}}}
\put(0,0.0){{\psfig{figure=fig15c.ps,width=8.93cm,angle=-90.0}}}
\end{picture}
\caption[]{Variation of the observed color as a function of redshift $z$ for a 
template galaxy with the properties of \iz18. 
The model consists of a high-surface brightness {\nlx core} 
to within which stellar emission is confined and dominates 
{\bf (upper panel)}, and a larger lower-surface brightness nebular
{\nlx envelope} {\bf (middle panel)}. 
The colors of the {\nlx core} and the {\nlx envelope} yield for $z=0$ a good
match to the observed values for the core (\rr$\leq$6\arcsec) and the nebular
envelope (6\arcsec$\leq$\rr$\la$20\arcsec) of \iz18\ (cf Sect. \ref{phot:IZw18}).
The {\bf lower panel} shows the dependence of the color contrast \dc\ 
between the core and the envelope as a function of redshift.}
\label{IZw18-z}
\end{figure}

The variation of the $B$--$J$, $V$--$K$, $V$--$R$, $V$--$I$ and $R$--$I$ 
colors of the core and the envelope as a function of $z$ is shown in the 
upper two panels of Fig. \ref{IZw18-z}. It can be seen that the envelope (middle panel)
shows particularly large color variations, as different strong emission lines shift
into the transmittance window of various filters, depending on $z$.  
As already evident from Sect. \ref{phot:IZw18}, a local analog to \iz18\ is easily 
identifiable by its blue nebular envelope and large (0.8 mag) \dc\ (lower
panel) both with respect to $V$--$I$ and $R$--$I$.
This is also the case for the redshift range $0.15\la z \la 0.3$ where the envelope 
appears much redder than the core in $V$--$I$ but bluer in $B$--$V$.
Other distinct peaks in $\|$\dc$\|$ with respect to various optical or optical--NIR colors
are apparent for e.g. $z\approx 0.42$, 0.55 and 0.9.
A comparison of the upper and lower panel of Fig. \ref{IZw18-z} shows that
the \dc\ exhibits much larger variations ($\|$\dc$\|$$\simeq$1.6 mag) than the 
HSB core, i.e. it is a far more sensitive indicator of 
extended \ige\ than integral (luminosity-weighted) colors that are primarily
driven by the core.
For example, at $z\approx0.27$, the $V$--$I$ and $B$--$V$ colors of the
HSB core are equal to within $\leq$0.3 mag, whereas their \dc\ is differing 
by up to $\approx$1.3 mag.

It is worth pointing out that, for certain redshift windows and color indices, 
the \dc\ may point towards diametrically different views on the nature of an \iz18\ analog.
For instance, at $0.15\la z \la 0.3$ the large $V$--$I$
core-to-envelope color contrast (\dc\ $\sim$ 0.8 mag) and the moderately blue
colors of the core (0.5 mag) superficially suggest an old disk hosting nuclear SF activity.
The opposite conclusion could be drawn from the \dc\ in $B$--$V$
($\approx$0.5 mag) which may be taken as evidence for a very young 
stellar disk encompassing a slightly older core, in line with the 
inside-out galaxy growth interpretation.
In either case, the exponential envelope, usually interpreted as stellar disk, 
would nicely augment the erroneous conclusion. 
Several similar cases for various color combinations and redshifts can be read off
Fig. \ref{IZw18-z}.
Of importance is also the fact that for other $z$'s 
(e.g. 0.1$\la z \la$0.15, 0.3, 0.75, 1.05) the \dc\ vanishes to $\la$0.2 mag
for some colors, having little discriminating power between stellar 
and nebular emission, and superficially suggesting a nearly uniform stellar age. 

\smallskip
With regard to all concerns above, one might counter-argue that photometric 
$k$ corrections would rectify SBPs and color profiles, and eliminate pitfalls 
in the interpretation of \dc.
However, state-of-art $k$ corrections, mostly tailored to stellar SEDs and applied to a
morphological analog of \iz18\ as a whole, regardless of its physically 
distinct radial zones, would most probably aggravate the problem in a barely 
predictable manner.
%

\smallskip
In summary, the case of \iz18\ stands as a warning benchmark to studies of 
high-SSFR galaxies near and far.
It reminds us of the fact that nebular emission in SF galaxies is not 
always cospatial with the underlying local ionizing and non-ionizing 
stellar background, neither has to scale with its surface density. 
Quite the contrary, in galaxies with high SSFR (i.e. forming young 
galaxies and/or systems rapidly assembling their stellar mass during dominant
phases of their evolution) nebular emission is plausibly expected to extend 
far beyond the galaxy itself and to significantly contribute to its total luminosity.
This, in connection with the empirical evidence that nebular emission forms an
\emph{exponential} envelope, conspires in potentially important observational 
biases in studies of moderately resolved morphological analogs to \iz18\ 
which are likely ubiquitous in the early universe.

Interesting in this context is also the detection of \emph{exponential} Ly$\alpha$
  halos around low \citep{Hayes2007} and high-$z$ star-forming galaxies
    \citep{Steidel11-Lya}, resulting in SBPs strikingly similar to those of \iz18.

In principle, any strong thermal or non-thermal central source of energy and momentum could
generate a large nebular envelope and a pseudo bulge--disk luminosity profile.
The foregoing discussion is therefore of relevance to a wider range of topics, 
as e.g. the co-evolution of AGNs with their host galaxies or the properties of ionized
halos around powerful radio galaxies \citep[e.g.,][and references therein]{VillarMartin03}.

The prospect of spatially resolved studies of galaxy formation in the faraway universe
with next-generation observing facilities calls for theoretical guidance on the properties and time evolution 
of nebular halos around protogalactic systems.
Some questions (see also P02) to computational astrophysics include:
i) in which way are the photometric properties of the nebular envelope 
(e.g. its $\mu_0$ and $\alpha$), and their temporal evolution, related to the
specifics of energy production (e.g. the SFH and the ionizing SED), 
the initial geometry and kinematics 
of the protogalactic gas reservoir and the physical conditions in the intergalactic medium
(e.g. external pressure, ambient ionizing field from multiple mutually 'illuminating'
protogalactic units)? More specifically,
ii) does (for a given set of environmental conditions) the core-envelope
radial intensity pattern scale in a homologous manner, i.e. is the $\alpha$ 
of the nebular envelope invariant? If not, iii) what does the core-to-envelope
luminosity and normalized \rsf/$\alpha$ ratio tell us about e.g. the 
recent SFH and current SSFR? 
Additionally, iv) how is the photometric, chemical and kinematical evolution of the nebular
envelope related to the escape probability of Ly continuum and Ly$\alpha$ photons?

\section{Summary and conclusions \label{Conclusions}}
We used the entire set of archival \hst\ ACS broad band imaging data
for \iz18\ and its fainter component \cc\ to study the photometric structure
of this nearby (19 Mpc) dwarf galaxy pair to unprecedently faint surface
brightness levels ($\mu\ga$29 \sbb).

\smallskip
\noindent The main results from this study may be summarized as follows:

\noindent {\nlx i)} Radial color profiles reveal very blue and practically 
constant colors (0$\pm$0.05 mag) for \cc\ down to $\mu \sim 27.6$ \sbb, and 
a previously undisclosed, slightly redder ($V$--$I$=0.2$\pm$0.08 mag, 
$R$--$I$=0.16$\pm$0.09 mag) stellar component in its extreme periphery (27.6 -- 29 \sbb).
We have verified that these blue colors do not merely reflect the
luminosity-weighted average of blue young stellar clusters with a faint 
red stellar background but that they are characteristic for the unresolved
host galaxy of \cc.
We argue that the buildup of the photometrically dominant stellar component of
\cc\ has occurred in a largely sequential mode, through a star-forming (SF)
process that has likely started $\tau \sim $100 Myr ago at the redder 
northwestern tip of the galaxy and propagated with a mean velocity of 
$\sim$20 \kmsec\ to its bluer southeastern tip. 

The photometric properties of the extreme periphery of \cc\ are entirely consistent
with this formation scenario, if the effect of stellar diffusion is taken into account. 
Radial migration of newly forming stars with a mean velocity of $\approx$4 \kmsec\ over $\sim\tau$,
and the associated \emph{stellar mass filtering effect} described in 
Papaderos et al. (2002, hereafter P02), can naturally account for the 
slightly redder colors, absence of nebular emission (\ige) and topological 
properties of the stellar outskirts of \cc.
A faint ancient stellar substrate can not be ruled out, even though 
our analysis does not lend observational support to its presence.

\smallskip
\noindent {\nlx ii)} In \iz18\ (i.e. the main body) severe contamination 
of broad band colors by \ige\ prevents a conclusive age dating of 
the stellar component almost everywhere. Therefore, even though our combined images 
are the deepest presently available, we can not improve on the age-dating analysis
by P02 who have two-dimensionally subtracted strong nebular emission lines 
from broad band \hst\ WFPC2 images to isolate and age-date the residual underlying stellar
background. However, the colors of the reddest quartile of region $\omega$, a region at the
southeastern tip of \iz18\ with comparatively weak \ige,
were determined to be in good agreement with those previously inferred by P02.
These colors, if representative for the host galaxy of \iz18,
imply on the basis of a continuous or exponentially decreasing star formation
history (SFH) that \iz18\ has formed most of its stellar mass 
at a late cosmic epoch. This, together with our conclusions under {\nlx i)}, 
supports the view that both \iz18\ and \cc\ are cosmologically young objects
that have undergone a nearly synchronous evolution. 

\smallskip
\noindent {\nlx iii)} We show that $\sim$20\% of the isophotal
area of \iz18\ at 25 \sbb\ is severely affected by \ige, thus inaccessible 
to age dating studies via broad band colors and color-magnitude diagrams (CMDs).
The local impact of \ige\ manifests itself i.a. in a combination of
red ($\sim$0.5 mag) $V$--$R$ with extremely blue (--0.4 \dots --0.8 mag)
$V$--$I$ and $R$--$I$ colors.
Nebular emission shows considerable sub-structure, with numerous clumps and 
overlapping shells, and little spatial correlation with the local stellar
background. It thus can not be treated as a uniform foreground emitting layer 
and accurately subtracted out using standard point source photometry algorithms.
This likely results in substantial random and systematic errors that might not
be fully accounted for by the standard CMD error budget.
Further potential sources of systematic uncertainties stem from spatial
displacements (50 -- 100 pc) between the intensity maxima of the 
[O{\sc iii}]$\lambda\lambda$4959,5007 and \ha\ emission lines along ionized
gas shells. These may differentially affect the local background determination
in various filters, causing an artificial reddening of CMD point sources
in the interior of nebular shells (and {\sl vice versa}).

\smallskip
\noindent {\nlx iv)} Based on the extraordinarily deep combined \hst\ ACS images, we have
been able to study the \emph{exponential} low-surface brightness (LSB) envelope of
\iz18\ out to its extreme periphery using both surface brightness profiles
(SBPs) and color maps. 
These reveal uniform colors of $V$--$R$$\approx$0.55 mag, $V$--$I$$\approx$--1
mag and $R$--$I$$\approx$--1.4 mag all over the LSB component of \iz18,
corroborating the previous conclusion by P02 that this luminosity envelope 
is not due to a stellar disk but due to extended \ige.
Specifically, our analysis indicates that \ige\ dominates the line-of-sight intensity
beyond 3 effective radii (or, equivalently, for $\mu\geq$23.5 \sbb) and
extends as far out as $\sim$16 stellar exponential scale lengths.
The overall picture emerging from our analysis is therefore that 
\iz18\ is a cosmologically young object that consists of a compact,
high-surface brightness (HSB) core, within which stellar emission is confined
and dominates, and a much larger nebular LSB envelope.

\smallskip
\noindent {\nlx v)} We argue that the morphological properties of \iz18,
while unique among nearby SF dwarf galaxies, are probably typical among 
distant young galaxies in dominant phases of their 
evolution, during which they assemble their stellar component at high
specific star formation rates (SSFRs). 
The prodigious energetic output during such phases of rapid stellar mass
growth is expected to result into a large-scale gas ionization and acceleration, 
with \ige\ protruding much beyond the still compact stellar galaxy host.
These systems could thus bear strong morphological resemblance 
to \iz18, comprising a compact core that is dominated by stellar
emission and a much larger exponential nebular envelope.
A question of considerable interest is therefore, how the nebular envelope 
could impact photometric studies of moderately resolved morphological
analogs of \iz18\ at higher $z$.

\smallskip
A potential bias, already discussed in P02, arises from the fact that the nebular 
envelope mimicks due to its exponentiality and red $B$--$R$ color an evolved stellar disk.
Here, by using \iz18\ as a template, we extend previous considerations:

{\nlx v.1} We point out that the superposition of two exponential components 
of differing central surface brightness and scale length, approximating the
core and the envelope of a distant \iz18\ analog, may be barely distinguishable from a genuine 
S\'ersic profile with an exponent 2$\la\eta\la$5.
Therefore, S\'ersic models offer, in the specific context, a poor diagnostic for 
disentangling compact high-SSFR protogalaxies from massive galaxy spheroids.

{\nlx v.2} Nebular emission contributes at least 1/3 of the total luminosity 
of \iz18\ (P02 and this study). 
This luminosity fraction, if typical for its higher-$z$ analogs,
translates into a systematic error of $\ga$0.4 mag in all galaxy scaling
relations involving luminosities (e.g., the Tully-Fisher relation, and
relations between luminosity and metallicity, diameter, velocity dispersion).
Evidently, errors in total luminosities propagate into errors in galaxy mass
determinations via theoretical mass-to-light ratios or SED fitting techniques.
Moreover, since extended \ige\ can drastically affect a galaxy SBP 
as a whole, it also affects practically all secondary photometric
quantities that are derivable from it (e.g. the effective radius or various 
light concentration indices used for quantitative galaxy morphology studies).

{\nlx v.3} We investigate the variation of the color contrast \dc\ 
between the star-dominated core and the surrounding nebular envelope as a function of $z$.
This task is motivated by the fact that \dc\ provides a handy proxy 
to radial color gradients in galaxies and a valuable galaxy classification tool.
We show that for certain $z$ intervals, this quantity offers a powerful 
diagnostic for the identification of moderately resolved \iz18\ analogs.
Within other $z$ intervals, however, and depending on the color indices considered,
a superficial interpretation of \dc\ can further enhance galaxy
misclassification biases stemming from SBP fitting (cf v.1) and potentially 
impact our understanding of galaxy assembly over time.
State-of-art $k$ corrections applied to distant morphological analogs to
\iz18\ as a whole, i.e. regardless of their physically distinct radial zones, 
may aggravate observational and interpretation biases 
in a barely predictable manner.

\smallskip
In the era of spatially resolved studies of galaxy formation in the early
universe with next-generation observing facilities, a better theoretical 
understanding of the rise and fall of nebular galaxy halos over cosmic time
appears to be crucially important. 
In this respect, some questions to computational astrophysics are formulated.

\begin{acknowledgements}
Polychronis Papaderos is supported by a Ciencia 2008 contract, funded
by FCT/MCTES (Portugal) and POPH/FSE (EC). 
He also acknowledges support by the Wenner-Gren Foundation.
G\"oran \"Ostlin acknowledges support from the Swedish Research
council (VR) and the Swedish National Space Board (SNSB). 
He is a Royal Swedish Academy of Sciences research fellow, supported
from a grant from the Knut and Alice Wallenberg foundation. 
A careful reading and constructive report by the referee is very much appreciated.
This research has made use of data acquired with the 
European Southern Observatory telescopes, and obtained from 
the ESO/ST-ECF Science Archive Facility and of the NASA/IPAC 
Extragalactic Database (NED) which is operated by the 
Jet Propulsion Laboratory, CALTECH, under 
contract with the National Aeronautic and Space Administration.
\end{acknowledgements}
%

\end{document}